\documentclass[conference,compsoc]{IEEEtran}

\usepackage[available,functional,reproduced]{ieeebadges}
\usepackage{graphicx}
\graphicspath{{images/} {images/generic_chatbots/} {images/task_specific_chatbots/}}
\usepackage{xcolor}
\usepackage{soul}
\usepackage{multirow}
\usepackage{float}
\usepackage{booktabs}
\usepackage{hyperref}
\usepackage{xurl}
\usepackage{amssymb}
\usepackage{colortbl}
\usepackage{xcolor}
\usepackage{pifont}
\usepackage{tcolorbox}
\usepackage{enumitem}
\makeatletter
\newcommand{\copyrightfootnote}[1]{%
  \begingroup
  \long\def\@makefntext##1{\noindent##1}%
  \footnotetext{#1}%
  \endgroup}
\makeatother

\usepackage{textcomp}
\newcommand{\circled}[1]{\textcircled{\raisebox{-0.3pt}{\small #1}}}
\newcommand{\etal}{et al.}

\begin{document}

\title{Penny for Your Prompts: A Measurement Study\\of Contextual Leakage in AI Chat Platforms}

\author{\IEEEauthorblockN{Paschalis Bekos}
\IEEEauthorblockA{Stony Brook University\\
Stony Brook, NY, USA\\
pbekos@cs.stonybrook.edu}
\and
\IEEEauthorblockN{Billy Tsouvalas}
\IEEEauthorblockA{Stony Brook University\\
Stony Brook, NY, USA\\
vtsouvalas@cs.stonybrook.edu}
\and
\IEEEauthorblockN{Nick Nikiforakis}
\IEEEauthorblockA{Stony Brook University\\
Stony Brook, NY, USA\\
nick@cs.stonybrook.edu}
\and
\IEEEauthorblockN{Michalis Polychronakis}
\IEEEauthorblockA{Stony Brook University\\
Stony Brook, NY, USA\\
mikepo@cs.stonybrook.edu}}

\maketitle

\begin{abstract} AI chatbot platforms have become one of the most popular
    web interfaces users interact with online, routinely receiving personal and intimate
    information shared under an expectation of privacy. Yet, on the web application side,
    these platforms have the ability to embed
    third-party analytics, advertising, and session replay scripts that operate
    silently alongside the chatbot, creating an underexamined channel for
    conversational data leakage towards third parties. We conduct the first
    systematic large-scale measurement study of contextual leakage in AI chatbots,
    auditing the web interfaces of 42 general-purpose and
    domain-specific (health, legal, and education) platforms. Beyond standard network
    interception, we instrument the browser at the JavaScript engine level to
    attribute each data flow to the exact user action that triggered it, and the DOM
    capture mechanism used to retrieve its content---granularity invisible to
    network-layer analysis alone. We find that 21 of these platforms leak conversation titles to third-party trackers,
    while nine leak verbatim user prompts. Beyond content, we observe a substantial amount of chatbot platforms
    sending user emails to third-party trackers, enabling the association of conversational context with a user's identity.
    Our findings cast traditional web privacy issues under a new light, highlighting the privacy issues
    that are inherent in how users interact with modern AI chatbots.
\end{abstract}

\copyrightfootnote{\copyright~2027 IEEE. Personal use of this material is permitted. Permission from IEEE must be obtained for all other uses, in any current or future media, including reprinting/republishing this material for advertising or promotional purposes, creating new collective works, for resale or redistribution to servers or lists, or reuse of any copyrighted component of this work in other works.}
\IEEEpeerreviewmaketitle

\newcommand{\statGmailTotalReq}{528}
\newcommand{\statGmailFirstReq}{266}
\newcommand{\statGmailThirdRaw}{262}
\newcommand{\statGmailThirdTracker}{135}
\newcommand{\statGmailBlocked}{12}
\newcommand{\statGmailUniqRaw}{10}
\newcommand{\statGmailUniqTracker}{7}
\newcommand{\statGmailNPrompts}{54}
\newcommand{\statGmailPctTracker}{24}

\newcommand{\statOtherTotalReq}{541}
\newcommand{\statOtherFirstReq}{336}
\newcommand{\statOtherThirdRaw}{204}
\newcommand{\statOtherThirdTracker}{107}
\newcommand{\statOtherBlocked}{10}
\newcommand{\statOtherUniqRaw}{10}
\newcommand{\statOtherUniqTracker}{7}
\newcommand{\statOtherNPrompts}{38}
\newcommand{\statOtherPctTracker}{19}

\newcommand{\statAllTotalReq}{540}
\newcommand{\statAllFirstReq}{295}
\newcommand{\statAllThirdRaw}{239}
\newcommand{\statAllThirdTracker}{118}
\newcommand{\statAllBlocked}{11}
\newcommand{\statAllUniqRaw}{10}
\newcommand{\statAllUniqTracker}{7}
\newcommand{\statAllNPrompts}{92}
\newcommand{\statAllPctTracker}{22}

\newcommand{\meantrackersession}{14}
\newcommand{\mediantrackersession}{8}
\newcommand{\uniquetrackerdomains}{153}
\newcommand{\meantrackersessionround}{13}

\newcommand{\eduTotalRequests}{514}
\newcommand{\healthTotalRequests}{169}
\newcommand{\legalTotalRequests}{393}
\newcommand{\allTotalRequests}{353}

\newcommand{\eduFirstPartyRequests}{185}
\newcommand{\healthFirstPartyRequests}{66}
\newcommand{\legalFirstPartyRequests}{176}
\newcommand{\allFirstPartyRequests}{136.78}

\newcommand{\eduThirdPartyTrackerRequests}{198}
\newcommand{\healthThirdPartyTrackerRequests}{70}
\newcommand{\legalThirdPartyTrackerRequests}{162}
\newcommand{\allThirdPartyTrackerRequests}{140}

\newcommand{\eduTrackerPct}{38.5}
\newcommand{\healthTrackerPct}{41.6}
\newcommand{\legalTrackerPct}{41.1}
\newcommand{\allTrackerPct}{39.7}

\newcommand{\totalTrackerRequests}{5{,}046}
\newcommand{\distinctTrackers}{106}

\section{Introduction}
\label{section:introduction}

AI chatbot platforms have become one of the most widely used web-based services, with hundreds of millions of users submitting queries that
range from general information requests to intimate personal
disclosures~\cite{chatbot-statistics-1, chatbot-statistics-2,chatbot-statistics-3,chatbot-statistics-4}.
Unlike traditional
search engines, where queries are typically short and do not encapsulate content in detail, chatbot interactions are typically far more detailed and contextual: users describe
medical symptoms, seek legal advice, disclose personal circumstances, and maintain extended conversations that accumulate sensitive context over time.
This shift fundamentally raises the privacy stakes of web activity, as chatbot interactions often contain substantial sensitive personal information.

Yet, from a web-infrastructure standpoint, AI chatbot platforms are built and operated like any other commercial web application, embedding the same third-party scripts found across the broader web. Among these authorized third parties are the same tracking
pixels and analytics frameworks that have been extensively studied in the context of traditional websites~\cite{fbclid-paper,inv-pixels,cookie-swap,cookie-graph,saly,children-websites-tracking,sick-and-tracked,k12-schools,religious,phi-leakage,medical}.
These frameworks are designed to monitor user interactions, attribute actions to conversions, and infer interests from the content a user engages with. On a
traditional website, this amounts to observing which products users browsed or which articles they read.
On an AI chatbot, however, the same mechanisms operate over content that is not passively consumed but \emph{actively} disclosed by the user, potentially
carrying medical, legal, financial, or deeply personal context---a distinction
that has received limited attention.

The gravity of this issue is underscored by class-action complaints recently filed against OpenAI~\cite{lim-v-openai,couture-v-openai} and
Perplexity~\cite{doe-v-perplexity}, alleging that embedded trackers
transmitted conversational context to third parties upon user interaction with
these platforms.
The research community has only begun to surface the problem. Jazlan~\etal~\cite{jazlan2026tracking} audited 20 general-purpose
chatbots using single-prompt network capture, and another study~\cite{leakylm}
examined four prominent platforms through HAR (HTTP Archive) traffic analysis. Both document the presence of third-party trackers and
instances of data leakage, but at a limited scale and depth.
First, they neglect the interaction-driven nature of tracking pixels
and miss the observation of additional leaks that only occur after multiple
interactions, such as the leakage of conversation titles that are generated
after the first user prompt.
Second, they focus primarily on general-purpose chatbots, while there is a
growing ecosystem of task-specific chatbots in sensitive domains such as health,
legal, and education. These platforms are likely to collect even more sensitive
context, yet they have been largely overlooked in prior research. As a result,
the scope and mechanics of contextual leakage in these platforms remain largely
unexplored.

In this work, we present a systematic large-scale measurement study of \emph{contextual leakage} in AI chatbot platforms. We audit the web interfaces
of 42 platforms, including 25 general-purpose chatbots and 17 task-specific
agents in sensitive domains (health, legal, and education).
Our study combines Puppeteer-based network capture with
VisibleV8~\cite{vv8} JavaScript (JS) engine instrumentation to attribute each data
flow to its causal user interaction and context extraction mechanism. Using Puppeteer, we conduct structured multi-interaction
sessions across each platform, capturing all outbound network traffic to identify third-party recipients of conversational
context and user identity. We then employ VV8 to isolate five interaction scenarios per platform, attributing each leakage flow
to its causal trigger. Beyond interaction attribution, VV8's execution traces allow us to reconstruct the JS-level mechanisms
through which third-party scripts retrieve conversational context from the structural components of the web page.

Our study allows us to identify conversational context leakage across AI chatbot platforms. Specifically, we find that 21 platforms
leak conversation titles and 9 leak verbatim user prompts to third-party
trackers.
Moreover, in the general-purpose chatbot platforms
we identify that 11 of the 13 context-leaking platforms also expose the user's email to trackers, enabling linkage of conversational
context to a named individual. Our analysis of the JavaScript execution traces reveals that the mechanisms deployed by
third-party scripts to retrieve descriptive page content on traditional websites, inadvertently capture conversational context on AI chatbot platforms. We find that each chat session is treated as a separate sub-page within the platform and assigned a
dynamically generated title derived from the user's prompt, which these scripts then read and transmit as if it were ordinary
page metadata.
Our key contributions are:
\begin{itemize}
      \item We audit 42 AI chatbot platforms, covering both general-purpose and task-specific chatbots in sensitive domains
            (health, legal, and education).
      \item We identify conversational context leakage alongside identity leakage towards third-party entities,
            revealing that trackers receiving conversation context can also link it to the identity of the user.
      \item We attribute each leakage flow to its causal user interaction and reconstruct the JS-level mechanisms
            trackers use to extract conversational context from the page.
      \item We show that third-party scripts deployed on AI chatbot platforms inadvertently capture conversational
            context through the same mechanisms they use on traditional websites, as the dynamic nature of chatbot interfaces
            causes page-level attributes to encode user intent rather than generic page metadata.
\end{itemize}

\section{Background}

\begin{figure}[t]
    \centering
    \includegraphics[width=\linewidth]{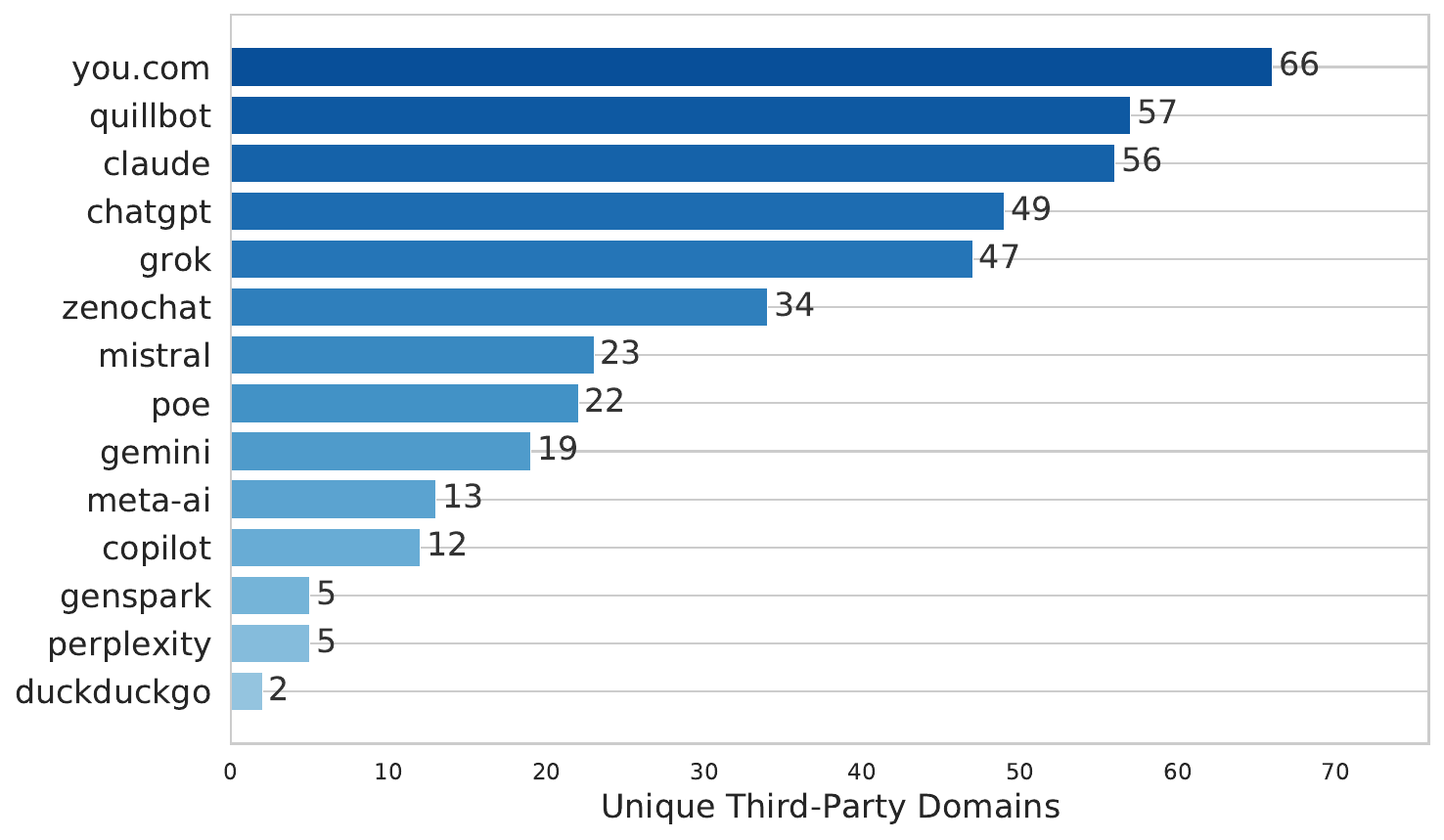}
    \caption{Number of unique third-party domains referenced in the Content Security
        Policy (CSP) of each general-purpose AI chatbot.}
    \label{fig:csp_third_party_domains_generic}
\end{figure}

The preliminary analysis of the Content Security Policies (CSPs) of some popular general-purpose chatbot platforms, as illustrated in Figure~\ref{fig:csp_third_party_domains_generic}, reveal a dense third-party ecosystem. Platforms such as
You.com, QuillBot, and Claude reference as many as 66, 57, and 56 distinct third-party domains, respectively, in their CSP headers alone. While a CSP entry does
not necessarily imply active data collection, it authorizes third-party scripts to execute within the platform's security context, establishing the
potential for data flows between the chatbot interface and external entities. This observation motivates a deeper investigation into how these third parties operate in the context of a
user's navigation and use of these platforms, and crucially, what data they are able to collect.

\subsection{AI Chatbots}

AI chatbots are web-based or application-based conversational interfaces powered by
large language models (LLMs) that simulate human conversation and generate
natural language responses to user queries. In recent years, users have
integrated these platforms into everyday use, submitting prompts
ranging from generic questions and work-related inquiries to sensitive personal
topics. Platforms such as ChatGPT, Gemini, and Claude have seen rapid consumer
adoption~\cite{chatbot-statistics-1,chatbot-statistics-2,chatbot-statistics-3,chatbot-statistics-4},
serving hundreds of millions of users across a wide range of
tasks, including information retrieval, content generation, and personal
assistance.

A distinguishing characteristic of AI chatbots compared to traditional web
applications, websites, or search engines, is the inherently contextual and
potentially sensitive nature of user input. Unlike a typical search engine
query, which usually consists of a few keywords, a chatbot prompt often contains
detailed disclosures ranging from generic topics (research questions, open-ended
inquiries, task assistance) to sensitive ones (medical symptoms, legal
situations, financial concerns, or personal inquiries). A further implication
arises from the preservation of \textit{history} across chats, which maintains
both the context and memory of prior questions. To that end, most platforms
automatically generate conversation titles that summarize the user's inquiry,
creating a second, platform-derived artifact that encodes the semantic context
of the interaction.

\begin{figure}[t]
    \centering
    \includegraphics[width=\linewidth]{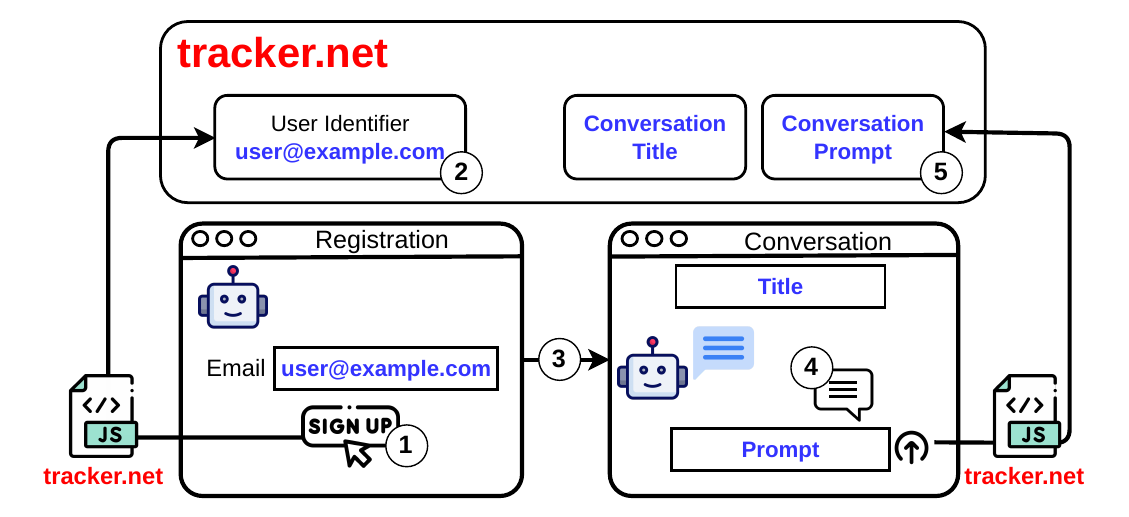}
    \caption{Overview of contextual leakage from an AI chatbot platform to a third-party tracker.}
    \label{fig:background-image}
\end{figure}

Figure~\ref{fig:background-image} illustrates how conversational context can be exposed to third-party entities alongside personally identifiable information (PII),
enabling the association of a user's identity with their conversational context. To illustrate the leakage model, we consider a scenario involving three actors:
a user~X, a web-based AI chatbot platform~C, and a third-party tracker~T operating under the domain \texttt{tracker.net}. We assume that T has no prior
information linking user~X to any online identity.

The user first visits the AI chatbot platform~C~\circled{1}. Upon page load, the platform includes a third-party tracking script served by~T. The user proceeds to create
an account by filling out and submitting a registration form~\circled{2}. This submission event is captured by an event listener registered by T's
script, which extracts the user's email address (\texttt{user\_email}) from the form and transmits it to T's server~\circled{3}. At this point, T can
associate a persistent identifier with user~X. The user then navigates to a new chat window~\circled{4} and submits a prompt containing a personal
inquiry. The platform processes the query, generates a response, and dynamically assigns a \emph{title} to the conversation that summarizes the user's
input~\circled{5}. This title is rendered into the page's DOM, typically as
the \texttt{<title>} element of the page or within a sidebar navigation entry. A
subsequent
browser event triggers T's tracking script to read the page state again. The
script now accesses the conversation title and, depending on the platform,
the prompt text itself, and transmits this contextual information to T's server~\circled{6}. From this point onward, tracker~T possesses both the user's email
address and the semantic context of their conversation. By joining these two data points, T can associate a previously anonymous user with the substance
of their private inquiry, effectively \emph{deanonymizing} the conversational context.

This type of contextual leakage, particularly when coupled with identifiers that link conversation context to an identifiable user,
has not received sufficient attention. Recent work has studied data collection
within the GPT app ecosystems~\cite{gpt-specific-1,gpt-specific-2}, while a
recent report~\cite{koi-report} identified malicious extensions exfiltrating conversations across 10 AI platforms. Jazlan \etal~\cite{jazlan2026tracking}
presented the first measurement of third-party tracking on chatbot web platforms, finding a subset of chatbots sharing contextual information
with at least one third party. However, to date, no study has traced the full leakage chain from user interaction to JavaScript-level extraction mechanism,
nor has extended the analysis to task-specific chatbots in sensitive domains.

\subsection{Tracking Pixels}

The online activity of a user is a rich source of information that can be leveraged to understand behavior, tendencies, and preferences. Website operators
routinely embed third-party tracking scripts, commonly referred to as \emph{tracking pixels}~\cite{fbclid-paper,inv-pixels,cookie-swap,cookie-graph,saly}, to support analytics, advertising attribution, and conversion measurement.
These JavaScript-based trackers have been widely used across the web and are deployed by platforms such as Meta, Google, and TikTok to enable targeted advertising
at scale. When deployed on a website, these pixels can read contextual signals
from the webpage, including the URL, page title, and referrer, and
propagate the
collected data to their own servers via network requests typically embedded as values to the request's payload. Beyond page-level context, many trackers register event
listeners on forms and interactive elements, enabling them to capture user-provided PII, such as email addresses, names, and phone numbers.

Prior work has extensively documented the privacy implications of tracking scripts,
revealing their widespread use across the web (including in sensitive categories of
websites)~\cite{children-websites-tracking, sick-and-tracked,k12-schools,
    religious,phi-leakage,medical}, studying their data collection from a website's
context~\cite{piixel-leaks,passive-3}, and their ability to collect
PII~\cite{pii-leakage-1, pii-leakage-2, pii-leakage-3, pii-leakage-4,
    contact-forms-leakage, leaky-forms, registration-forms-large-scale,
    not-register, kieserman2025tracker}.  The adoption of tracking scripts by AI
chatbots introduces additional privacy implications, as these platforms
render user-generated content directly into the DOM. Prompts, responses, and
dynamically generated conversation titles all become part of the page structure,
readable by any executing script. The same DOM read that captures a news
headline on a publisher's site can capture a summary of a user's private query on a
chatbot. The exposed context is therefore not just \emph{who} the user is, but
\emph{what} they are asking about, seeking help with, or disclosing in
confidence.

\section{Methodology}

To identify cases of contextual leakage from AI chatbot platforms to third-party
trackers, we design a systematic measurement methodology that combines network
capture with JavaScript-level instrumentation.
Our approach consists of two main complementary phases:
\emph{session capture} and
\emph{interaction decomposition}.
For session capture, we instrument a Puppeteer-driven
browser that allows us to manually interact with each platform across structured sessions, capturing all outgoing traffic.
This enables the identification of third-party data collection of conversation context and personally identifiable information (PII).
While network interception reveals the full scope of data collection, it cannot attribute a given data flow to a specific user
action or reveal \emph{how} trackers extract conversation context from the
page.
To address this, we implemented an interaction decomposition approach
based on VisibleV8 (VV8)~\cite{vv8},
an instrumented build of the Chromium browser that logs all JavaScript API calls at the V8 engine level, enabling us to decompose each leakage
event across its causal trigger and the underlying extraction mechanism.
We use this methodology to audit a wide range of general-purpose and
task-specific AI chatbots.

\subsection{AI Chat Platform Selection}

Our study focuses on AI chatbots that provide a web-based interface, as users typically interact with these platforms through a standard browser session.
We selected a total of 42 chatbot platforms; 25 general-purpose and 17 task-specific ones. For the general-purpose platforms, our selection was based on publicly-available popularity rankings~\cite{similar-web}. The task-specific chatbots were selected to span educational, health-oriented, and legal-assistance platforms; three domains explicitly chosen because of the inherent sensitivity of the conversations users share within them.

Table~\ref{tab:platforms} lists the platforms included in our study. Although several general-purpose platforms target a specific niche,
such as, creative writing assistance (Writesonic, QuillBot), image generation (SeaArt, Genspark), or character-based roleplay (Character.AI, PolyBuzz),
all selected platforms expose a general conversational interface that accepts arbitrary user input, distinguishing them from task-specific
tools that restrict queries to a narrow domain. The task-specific platforms fall under the sensitive category with respect to their
primary conversational target, covering (i)~\emph{Education}, (ii)~\emph{Health}, and (iii)~\emph{Legal} assistance. In the case of iAsk we consider it for two categories:
(i) Educational, which is the primary category, and (ii) Health, as it offers iAsk companion-mode, a virtual companion focused on emotional and mental well-being support.

\begin{table}[t]
  \centering
  \caption{AI chatbot platforms included in our study.}
  \label{tab:platforms}
  \scriptsize
  \setlength{\tabcolsep}{3pt}
  \begin{tabular}{lll}
    \toprule
    \multicolumn{3}{c}{\textbf{General Purpose}}                                                                          \\
    \midrule
    Microsoft Copilot~\cite{copilot}                     & DeepSeek~\cite{deepseek}           & Claude~\cite{claude}      \\
    Mistral Le Chat~\cite{mistrallechat}                 & ZenoChat~\cite{zenochat}           & Meta AI~\cite{metaai}     \\
    Character.AI~\cite{characterai}                      & QuillBot~\cite{quilbot}            & Manus~\cite{manus}        \\
    HuggingChat~\cite{huggingchat}                       & Qwen Chat~\cite{qwen}              & Grok~\cite{grok}          \\
    Writesonic~\cite{chatsonic}                          & You.com~\cite{you}                 & Kimi~\cite{kimi}          \\
    Perplexity~\cite{perplexity}                         & ChatGPT~\cite{chatgpt}             & Poe~\cite{poe}            \\
    DuckDuckGo~\cite{duckduckgo}                         & SeaArt~\cite{seart}                & Pi~\cite{pi}              \\
    Genspark~\cite{genspark}                             & ChatOn~\cite{chaton}               &                           \\
    PolyBuzz~\cite{polybuzz}                             & Gemini~\cite{gemini}               &                           \\
    \midrule
    \multicolumn{3}{c}{\textbf{Education}}                                                                                \\
    \midrule
    Khanmigo~\cite{khanmigo}                             & EduBrain~\cite{edubrain}           & Flexi~\cite{flexi}        \\
    SciSpace~\cite{scispace}                             & Elicit~\cite{elicit}               & iAsk~\cite{iaskai}        \\
    Answer.AI~\cite{answerai}                            &                                    &                           \\
    \midrule
    \multicolumn{3}{c}{\textbf{Health}}                                                                                   \\
    \midrule
    Free AI (not a) Therapist~\cite{freeainotatherapist} & FeelBetterBot~\cite{feelbetterbot} & Wellzy~\cite{wellzy}      \\
    iAsk (companion mode)~\cite{iaskai-companion}        & Earkick~\cite{earkick}             & Abby~\cite{abby}          \\
    AITherapistFree~\cite{aitherapistfree}               &                                    &                           \\
    \midrule
    \multicolumn{3}{c}{\textbf{Legal}}                                                                                    \\
    \midrule
    LawConnect~\cite{lawconnect}                         & NyayGuru~\cite{nyayguru}           & AI Lawyer~\cite{ailawyer} \\
    TheLawGPT~\cite{thelawgpt}                           &                                    &                           \\
    \bottomrule
  \end{tabular}
\end{table}

Our selection is limited by the requirement for free account subscription and the availability of a web-based interface that
accepts arbitrary natural language prompts. Starting from SimilarWeb's ``AI Chatbots and Tools'' ranking, we selected
platforms that expose a chat-based interaction model where users can submit free-form queries and receive AI-generated responses,
filtering out coding assistants, app-building environments, API routers, and enterprise platforms without free public access.
We focus exclusively on web-based interfaces and do not consider mobile applications, which may employ different
tracking configurations and require separate analysis. In total, we audited \emph{42 AI platforms} for both identity and contextual
leakage towards third-party entities. All measurements were conducted from a US East vantage point during May 6 and May 20, 2026.

\subsection{Session Capture}
\label{sec:capture}

We instrumented a Puppeteer-driven browser~\cite{puppeteer} to interact with each platform across multiple browsing sessions,
performing our analysis manually to remain faithful to the actual user experience. Each session followed a structured sequence of
user interactions:

\begin{enumerate}
  \item \emph{Account Creation}
  \item \emph{Prompt Submission:} We submit a prompt and observe
        the platform's response (``weather inquiry for a city~X'').
  \item \emph{Prompt Submission:} We submit a different prompt in
        a new chat window (``dentist recommendation for a
        neighborhood~Y in city~X'').
  \item \emph{Prompt Navigation:} Navigation between the two
        conversation windows.
  \item \emph{Session Termination}
\end{enumerate}

Each session is executed in a clean browser instance with no prior state, cookies, or cached data, and we interact with each platform
under its default configuration, accepting any cookie or consent prompts as presented without modifying chat settings or privacy
controls. We repeat each session under two separate accounts: one registered with a Gmail address and one with a custom email hosted
on our own mail server. This variation is applied only to platforms that support email-based registration. For platforms that allow interaction
without account creation, such as DuckDuckGo AI Chat, we capture a no-account baseline session.

Our prompts cover both a generic inquiry (weather) and a sensitive one (medical/dentist) and across all sessions we submit the same
exact prompts to enable consistent comparison. Each prompt yields a \emph{dynamically} generated title containing \emph{contextual}
information about the user's inquiry, which we record for each session. For the character-based platforms (Character.AI and PolyBuzz), 
we interact with the \emph{same} set of characters across sessions and record the resulting conversation title, which reveals the name of the 
character the user is conversing with. Throughout each session we capture all outgoing network
requests, recording the full URL, cookies, headers, and POST bodies. Each request is then analyzed for the presence of the
seeded prompt or the recorded title. We classify a leakage event when the domain receiving conversation context (the full prompt
or prompt title) or PII does not belong to the same parent domain (eTLD+1) or
an organizational domain from the same vendor (e.g., anthropic.com with
claude.ai). We further exclude CDNs, consent management platforms, payment processors and other infrastructure domains that do not
constitute tracking or advertising third parties.

For the task-specific chatbots, we follow the same session structure but limit registration to a Gmail account only, as the goal is to
measure whether sensitive conversation context is exposed to third parties rather than to re-evaluate the effect of registration method.
For each of the three categories, we submit two prompts: one generic and one carrying sensitive context reflective of the platform's
domain. Specifically, for \emph{education} platforms we submit a general academic reference request alongside a query seeking sources
on a sensitive personal topic; for \emph{legal} platforms we submit a general criminal law question alongside an immigration status
inquiry; and for \emph{health} platforms we submit a general wellness question alongside a query disclosing symptoms of anxiety and
interest in medication. Given that these platforms inherently handle sensitive user inquiries by nature of their domain, any contextual
leakage observed carries significantly higher privacy implications than that of general-purpose chatbots.

While this phase reveals third-party actors that collect conversation context, the multi-interaction session design cannot attribute which
specific user action triggered each observed data flow. A tracker receiving the conversation title during a session that included both
prompt submission and chat navigation could have been triggered by either action.

\subsection{Interaction Decomposition}

To address this attribution gap, we employ VisibleV8 (VV8), an instrumented build of the Chromium browser that logs all JavaScript
API calls at the V8 engine level, including property reads and writes, function invocations, and network-initiating operations.
VV8 produces execution traces that capture every interaction between JavaScript code and the browser's Web API surface, enabling us to
observe not only \emph{that} data was sent to a third party, but \emph{which} script read it, from \emph{which} DOM property, and
through \emph{which} network API.

\begin{figure}[t!]
  \centering
  \includegraphics[width=\columnwidth]{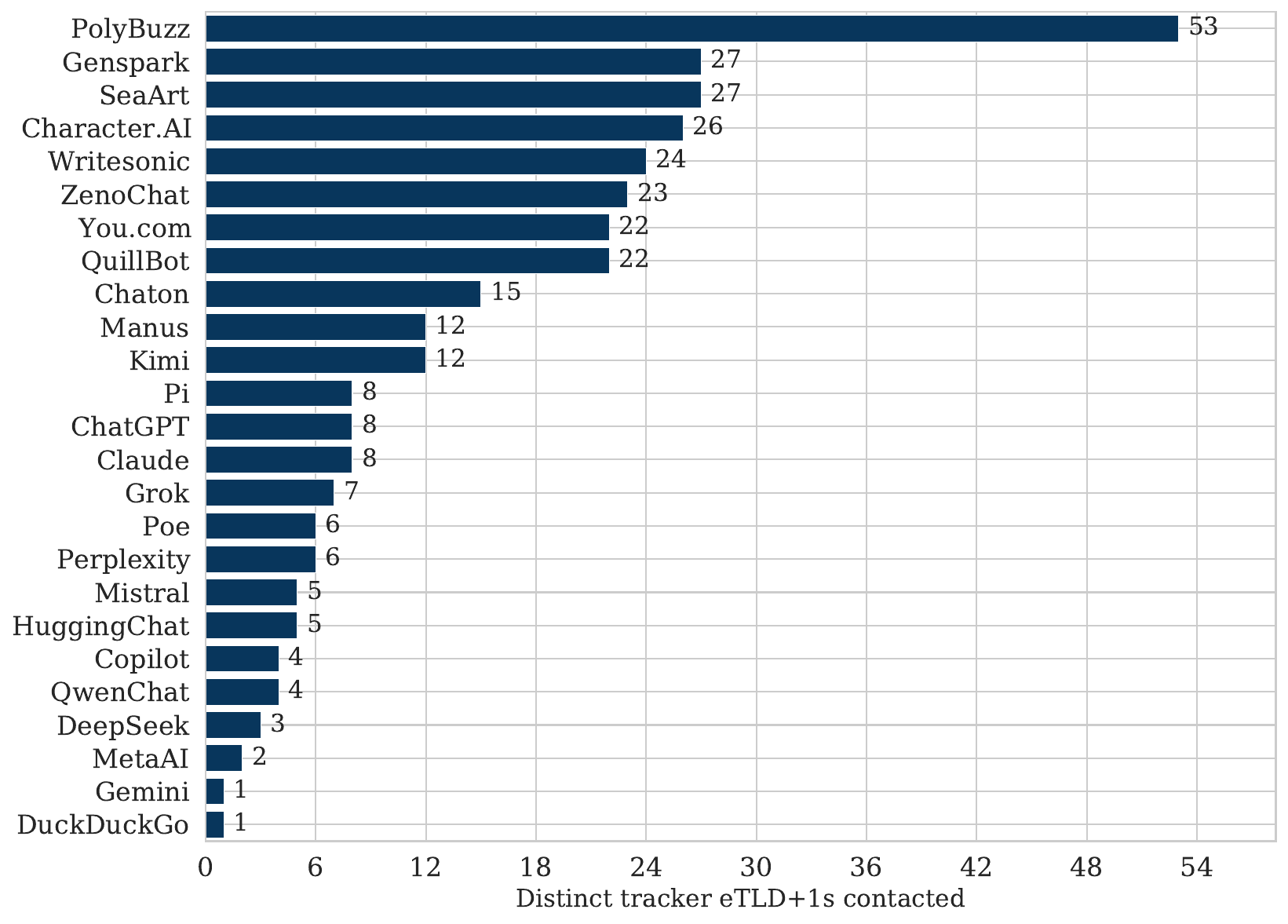}
  \caption{Unique tracker domains (eTLD+1) contacted by each chatbot.}
  \label{fig:unique-tracker-domains-per-chatbot}
\end{figure}

We design five isolated scenarios, each executed after login using the same email address in a separate VV8 session, producing independent
execution traces:
\begin{enumerate}
  \item \emph{Prompt Submission:} The user types a query, submits it, waits for the chatbot's response, and closes the browser.
        This establishes the baseline: what third-party collection occurs from the act of submitting a prompt alone.
  \item \emph{Chat Navigation:} The user submits a prompt, waits for the response, then navigates to a previously created
        conversation.
  \item \emph{Chat Deletion:} The user selects an existing conversation and deletes it, without submitting any new prompt.
  \item \emph{Scroll:} The user submits a prompt, waits for a sufficiently long response, and scrolls through it.
  \item \emph{Button Click:} The user submits a prompt, then clicks an unrelated UI element (such as a settings icon or
        sidebar button) that does not navigate to a different conversation.
\end{enumerate}

Each session is executed after the network is idle, and all scenarios use the same seed prompt (``Pharmacies near me CITY\_X, STATE\_Y'')
under the same browser configuration. For each scenario, we search all outbound network operations for two taint markers: the raw prompt and the
platform-generated conversation title, in their original form as well as URL-encoded variants. By comparing the third-party
destinations and leaked values observed in each scenario against the prompt-submission baseline, we determine which user action is
the causal trigger for each data flow.

Beyond this cross-platform differential analysis, VV8's execution traces allow us to reconstruct the full JavaScript execution chain leading to each observed leakage flow.
For every outbound request containing the prompt or title, we identify which script executed the request, resolve its origin, and enumerate DOM read
operations that the script performed before sending the data. This reveals the specific capture mechanism each tracker deploys to retrieve conversation context,
for example, a property access on \texttt{document.title} or a DOM query for the \texttt{<title>} element via \texttt{querySelector} or \texttt{getElementsByTagName}.
These JavaScript-level distinctions are invisible to network-layer capture, where all mechanisms produce identical HTTP request parameters, thus enabling
us to study not only \emph{what} context is leaked but \emph{how} it is extracted before propagation to a tracker.

\begin{figure}[t]
  \centering
  \includegraphics[width=\linewidth]{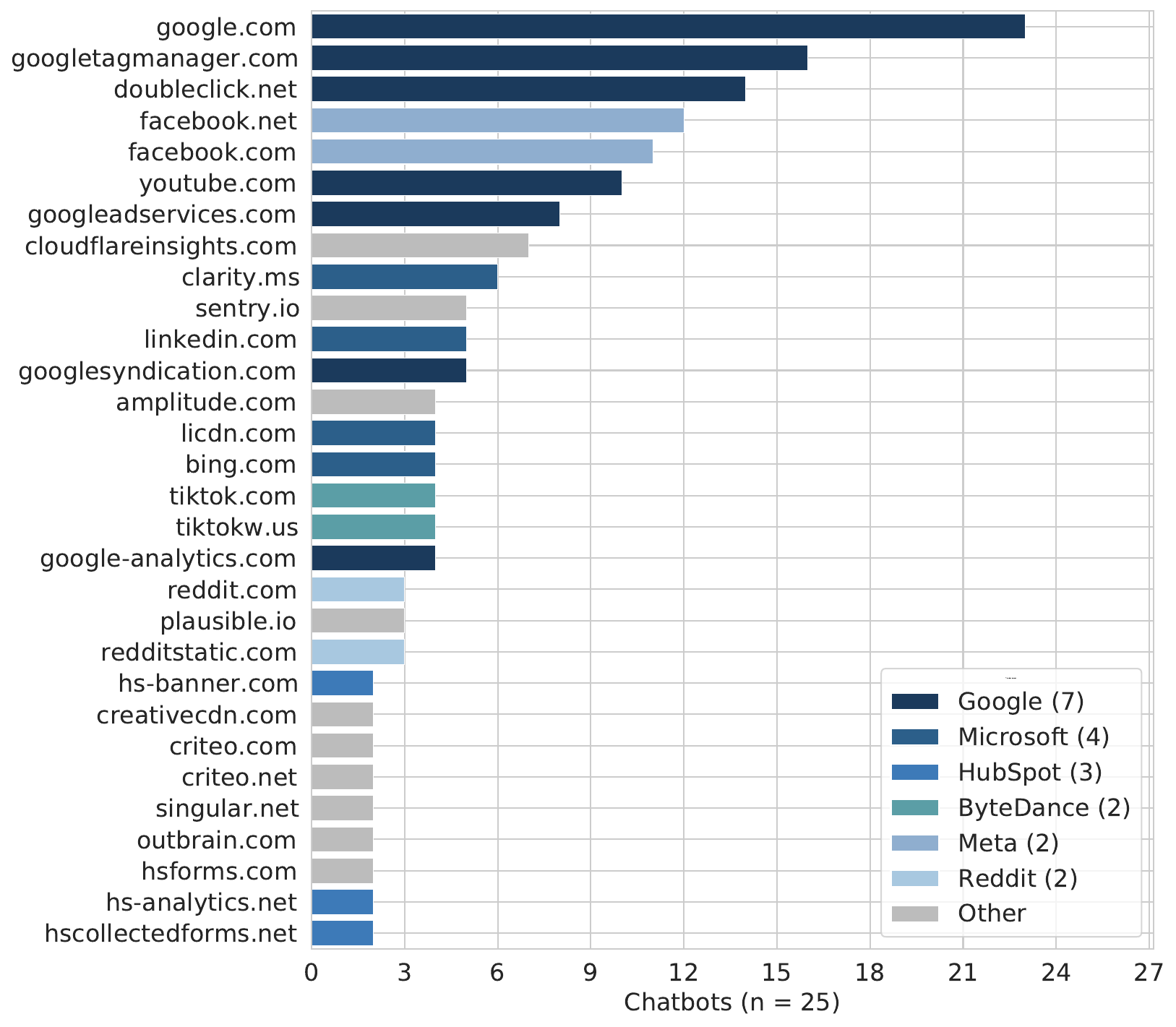}
  \caption{Third-party tracker prevalence across chatbots.}
  \label{fig:third_party_prevalence_across_the_chatbot_corpus}
\end{figure}

\section{Findings}
\label{section:findings}

In this section, we present the results of our measurement experiment across
42 AI chatbot platforms. Specifically, we characterize the prevalence and
volume of contextual leakage to trackers, identify the destinations receiving
conversation data, and examine how leakage patterns differ across account
types for general-purpose (25) and task-specific (17) AI chatbot platforms. Then, we attribute each leakage 
flow to the user interaction it originated from, and describe the underlying mechanism responsible for gathering and 
transmitting the data to third parties. Our per-prompt quantities reported throughout this section are averaged over the two 
seeded prompts submitted within each session.

We note that throughout our tracker analysis, we exclude domains related to
cloud infrastructure, CDNs, consent management, CAPTCHA, and any other non-tracker
service, as well as each platform's own related domains (e.g., \texttt{openai.com} for ChatGPT).
Both sets were curated manually, with non-obvious platform affiliations confirmed against vendor documentation (e.g., \texttt{oaistatic.com}
is listed on OpenAI's support pages). Remaining third-party eTLD+1s were attributed to their parent company from public corporate-ownership records
(e.g., \texttt{doubleclick.net} to Google, \texttt{clarity.ms} to Microsoft). We validated this attribution against Disconnect~\cite{disconnect-list} 
and DuckDuckGo Tracker Radar~\cite{tracker-radar}: 88.9\% of assignments are confirmed by both sources and 97.2\% by at least one, with no contradictions; the single unconfirmed 
domain (\texttt{gator.volces.com}) was attributed from vendor documentation. This ensures that all reported third-party flows reflect solely analytics, advertising, and behavioral-tracking services.

\begin{figure}[t!]
  \centering
  \includegraphics[width=\linewidth]{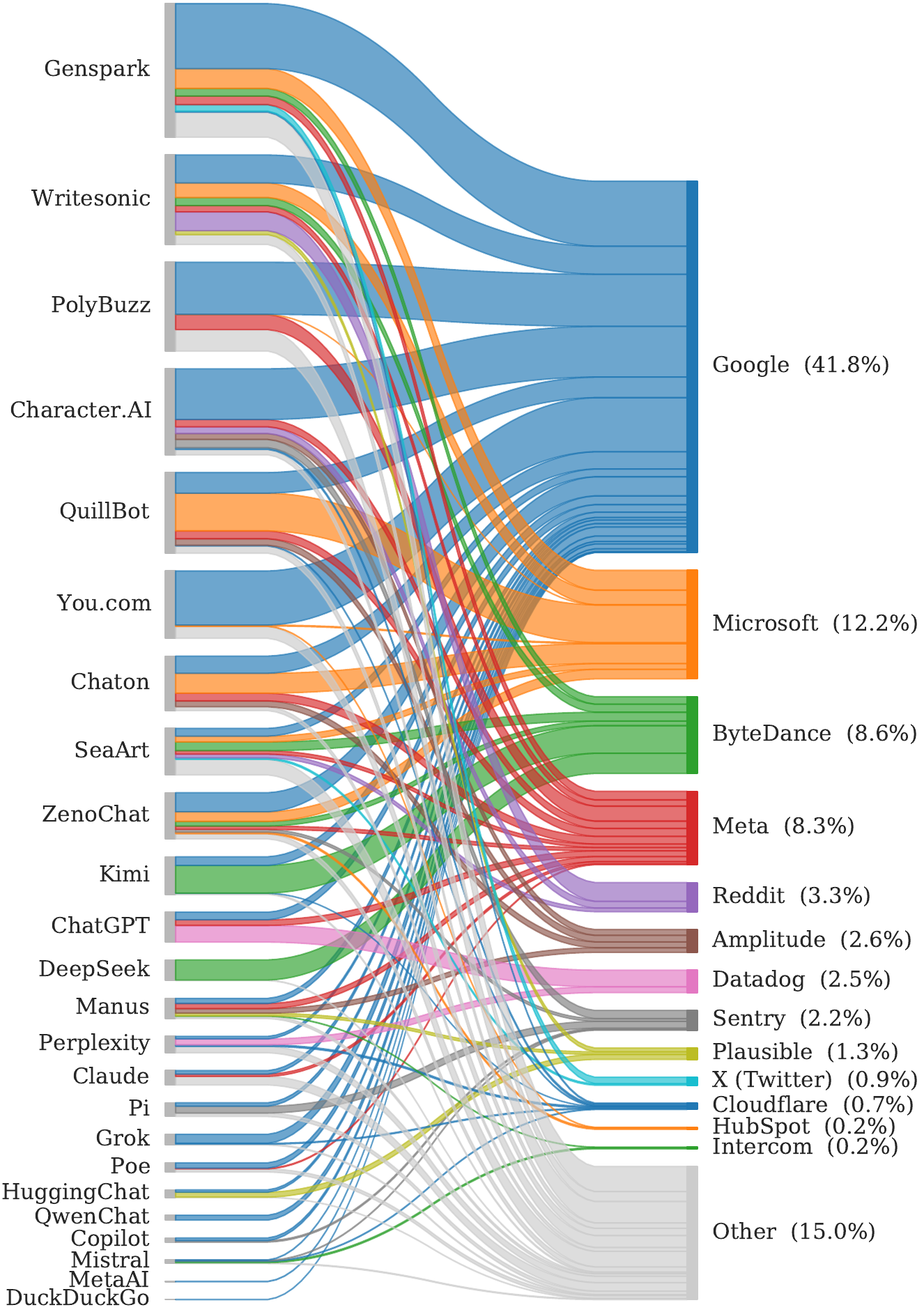}
  \caption{Third-party requests per generic chatbot platform to known tracker domains, aggregated by owner company.}
  \label{fig:top_30_3p_domains_x_chatbots_request_counts}
\end{figure}

\subsection{Trackers in Generic AI Chatbots}
\label{subsection:findings_trackers_generic_chatbots}

Following the steps outlined in Section~\ref{sec:capture}, we first analyze the
network traffic generated by each general-purpose chatbot platform.
On average, across the 25 general purpose chatbots, each prompt interaction generates \statAllTotalReq{} outbound
requests, of which \statAllThirdTracker{} (\statAllPctTracker\%) are directed to
known tracker domains. In terms of tracker domains, we identify
\uniquetrackerdomains{} unique trackers (eTLD+1) across all chatbots, with each
session involving \meantrackersessionround{} of them on average. In Figure~\ref{fig:unique-tracker-domains-per-chatbot}, we present the distribution of unique tracker domains across the general-purpose AI chatbots, showing how many distinct trackers each platform embeds.

\subsubsection{Tracker Traffic}
\label{subsubsec:findings_trackers_generic_chatbots_tracker_traffic}

Beyond their prevalence, we examine where tracker traffic ultimately concentrates. Since a single company operates many distinct domains, we consolidate them into \emph{owner blocks}, e.g., Google comprises \texttt{google.com, googletagmanager.com, doubleclick.net, googleadservices.com}, etc. Counting domains in isolation understates how much data converges on a single recipient, whereas owner blocks reveal the true entity ultimately receiving the requests. Applying this consolidation, in Figure~\ref{fig:third_party_prevalence_across_the_chatbot_corpus}, we present each tracker's reach across the corpus grouped by owner block, making clear how a few companies stand behind the many domains and appear on nearly every platform.

\begin{figure}[t!]
  \centering
  \includegraphics[width=\linewidth]{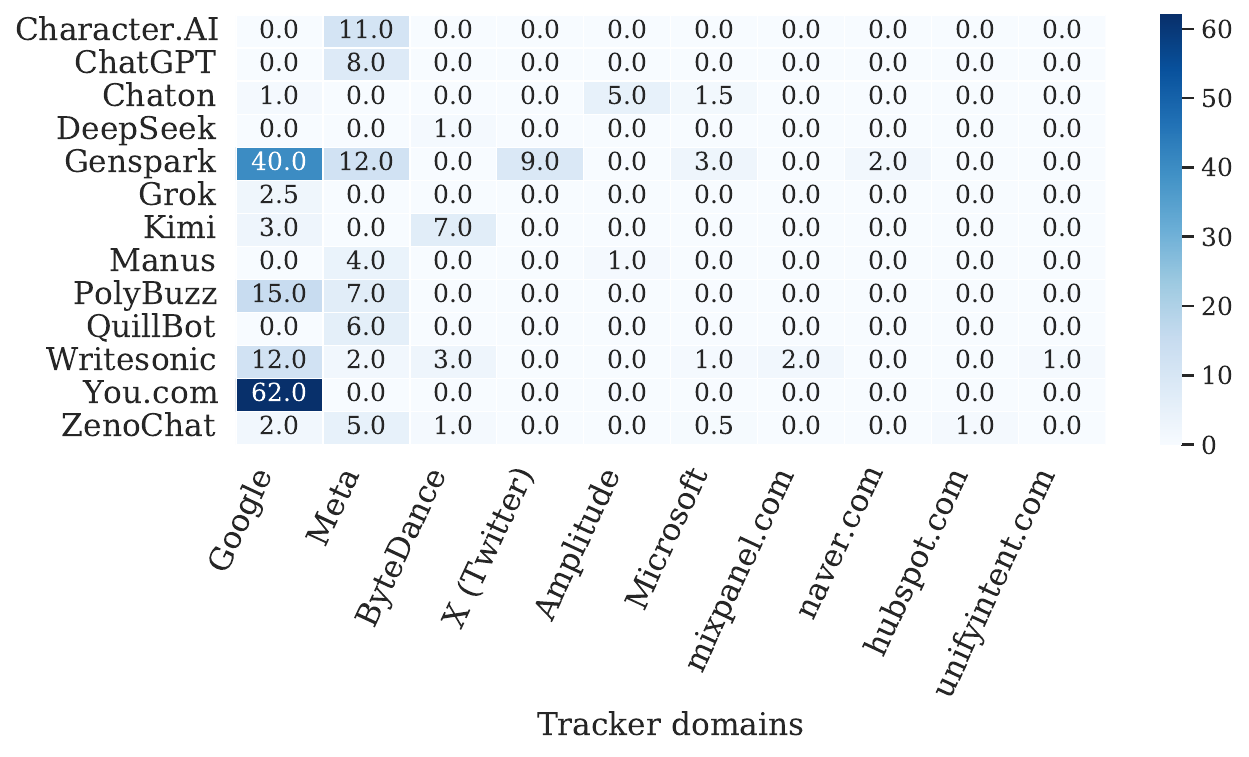}
  \caption{Whole-string title leaks per generic chatbot, averaged per prompt.}
  \label{fig:whole_string_title_leaks_chatbot_owner}
\end{figure}

Figure~\ref{fig:top_30_3p_domains_x_chatbots_request_counts} maps each of the
25 evaluated chatbot platforms to the tracker owner companies that receive
their outbound third-party requests.
Each ribbon connects a chatbot on the left to a tracker owner on the right, with ribbon width proportional to request volume. We observe that Google is the dominant recipient, accumulating more than 40\% of all tracker traffic, followed by Microsoft (12.2\%), ByteDance (8.6\%), and Meta (8.3\%).
Overall, \emph{more than 70\%} of all tracker traffic from the 25 chatbot platforms flows to just \emph{four companies}, providing them cross-platform visibility into user interactions across the AI chatbot ecosystem, regardless of which platform a user chooses.

\subsubsection{Conversational Context Leakage}
\label{subsubsec:findings_trackers_generic_chatbots_tracker_requests}

Having established that a significant share of outbound traffic is directed to
known trackers, we now examine these requests for conversational context, i.e.,
chat titles and raw user prompts. The heatmap of
Figure~\ref{fig:whole_string_title_leaks_chatbot_owner} presents the instances of
\emph{chat title leakage} per chatbot. These are cases where a chatbot makes a third-party request to a known 
tracker domain, containing the title of the conversation a user has with the chatbot.
Each cell corresponds to the average number of leaking requests per prompt
towards a domain owned by the respective vendor.
Overall, 13 of the 25 evaluated platforms send the title of a user's
conversation to a third party tracker. In terms of request volume, You.com and
Genspark lead this group, sending 62 and 40 title-bearing requests per prompt
to Google, respectively. Moreover, multiple chatbot platforms exhibit substantial leakage to more than one tracker destinations. 
These findings demonstrate that conversation title leakage is not an isolated phenomenon but a systemic one, with Google and 
Meta as the primary beneficiaries.

\begin{figure}[t!]
  \centering
  \includegraphics[width=\linewidth]{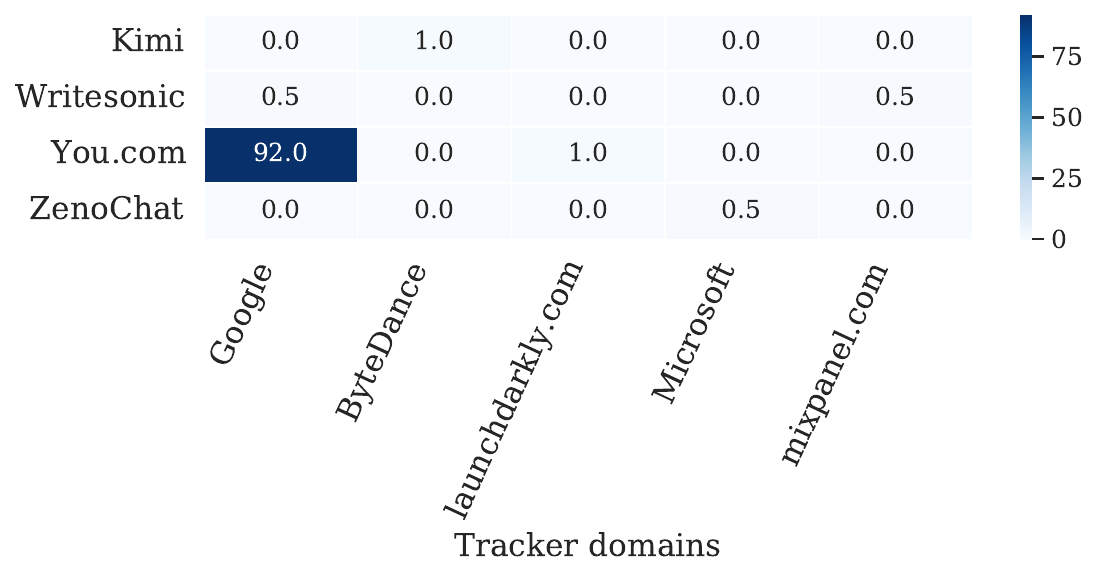}
  \caption{Whole-string prompt leaks per generic chatbot, averaged per prompt.}
  \label{fig:whole_string_prompt_leaks_chatbot_owner}
\end{figure}

We also identify direct \emph{raw prompt leakage} within the payload of
requests propagated to third-party entities, as shown in
Figure~\ref{fig:whole_string_prompt_leaks_chatbot_owner}. These are cases
where the verbatim user prompt appears in an outbound request
towards a known tracker domain. Four platforms exhibit this
more severe form of leakage, with You.com being the most prevalent, sending on average 92 prompt-bearing requests to Google every time a user
submits a prompt. Notably, Writesonic distributes the raw user prompt to three independent tracker endpoints. This is particularly concerning because
unlike conversation titles, prompts are composed entirely by the user and routinely contain sensitive information.

Finally, an interesting observation is the leakage asymmetry based on the email used to register an account. Accounts registered with 
a Gmail address exhibit higher tracker-bound request rates (\statGmailPctTracker\%) compared to accounts using other email providers 
(\statOtherPctTracker\%), and for most platforms the title or verbatim prompt reaches trackers more often under a Gmail account 
than a non-Gmail one, as we present in
Figure~\ref{fig:mail_provider_asymmetry_of_whole_string_leak_volume} in the Appendix.
However, with one 
session per configuration, confirming this asymmetry would require a future dedicated measurement.

\subsubsection{Context Leakage per Tracker}
\label{subsubsection:findings_trackers_generic_chatbots_}

Our per-platform analysis reveals which chatbot services leak conversational
context and to whom. To understand the aggregate exposure at the
ecosystem level, we evaluate which tracker-owning
vendors accumulate the most leaking requests.

\begin{figure}[t!]
  \centering
  \includegraphics[width=\linewidth]{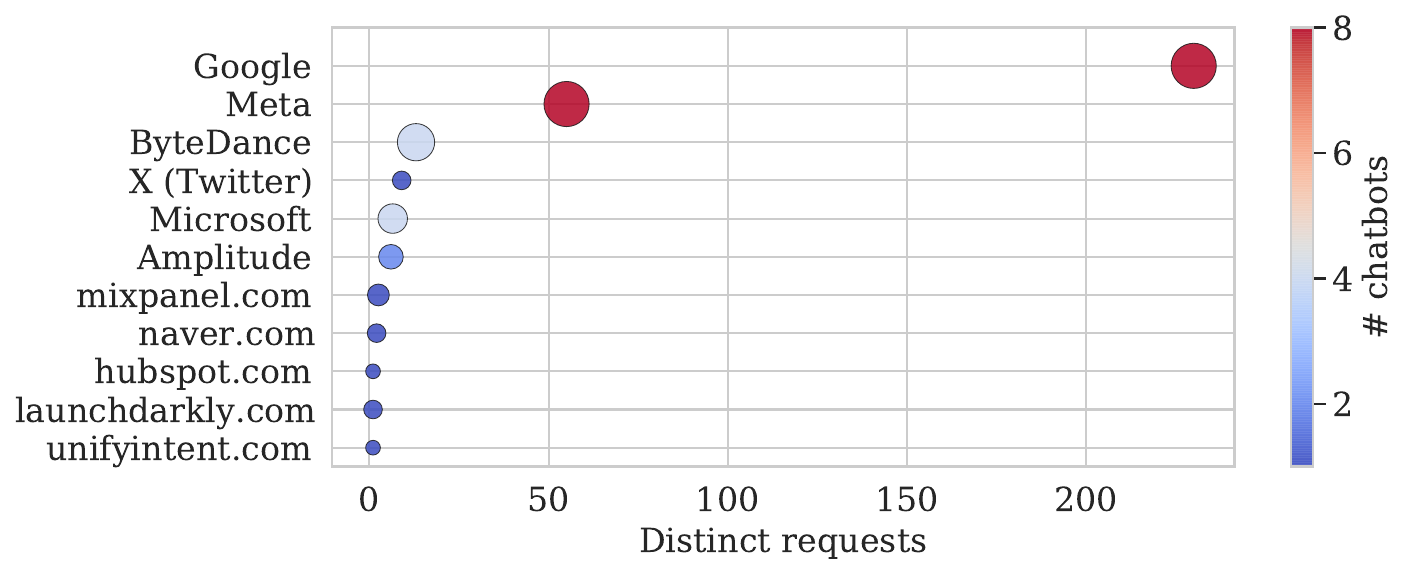}
  \caption{Distinct context-bearing (title or prompt) requests from generic chatbots to top tracker domains.}
  \label{fig:whole_string_leaks_top_3p_domains_distinct_requests_grouped_by_owners}
\end{figure}

Figure~\ref{fig:whole_string_leaks_top_3p_domains_distinct_requests_grouped_by_owners}
presents the number of \emph{whole-string title or prompt leakage}
requests received by each tracker-owning company.
Google accumulates 230 distinct context-bearing requests across eight chatbot platforms and Meta
follows with about 55 requests spanning the same number of platforms.  The
steep volume drop-off after Google and Meta, combined with their equal platform
breadth, indicates that conversational context leakage is structurally
concentrated in the two largest digital advertising entities, granting them
cross-platform visibility into what users discuss with AI chatbots.  ByteDance
ranks third, driven primarily by Kimi's title leakage, followed by X (Twitter)
and Microsoft, each receiving fewer than 15 distinct requests.
Figure~\ref{fig:whole_string_leaks_top_3p_domains_distinct_requests} in the
Appendix provides a detailed breakdown of whole-string title and prompt leakage by each
tracker destination domain.

\subsubsection{Email Leakage and Identity Correlation}
\label{subsubsection:findings_trackers_generic_identity}

So far, we have established \emph{what} leaks and to whom, but not whether any of it can be traced back to a real user. 
A leaked title or prompt rarely arrives alone: it is accompanied by the persistent identifiers the tracker has already placed on the user, 
most commonly a long-lived cookie (e.g. Google's \emph{NID} cookie). These identifiers allow the tracker-owning companies to stitch successive conversations into a 
single profile. Up to this point, this profile remains \emph{pseudonymous}, only referring to a unique cookie value, 
rather than a real-world identity. However, when the same tracker-owning company obtains a real-world identifier, 
e.g., the registration email, it is possible to associate the accumulated profile data to a real, named individual.

\begin{figure*}[t!]
  \centering
  \includegraphics[width=0.9\textwidth]{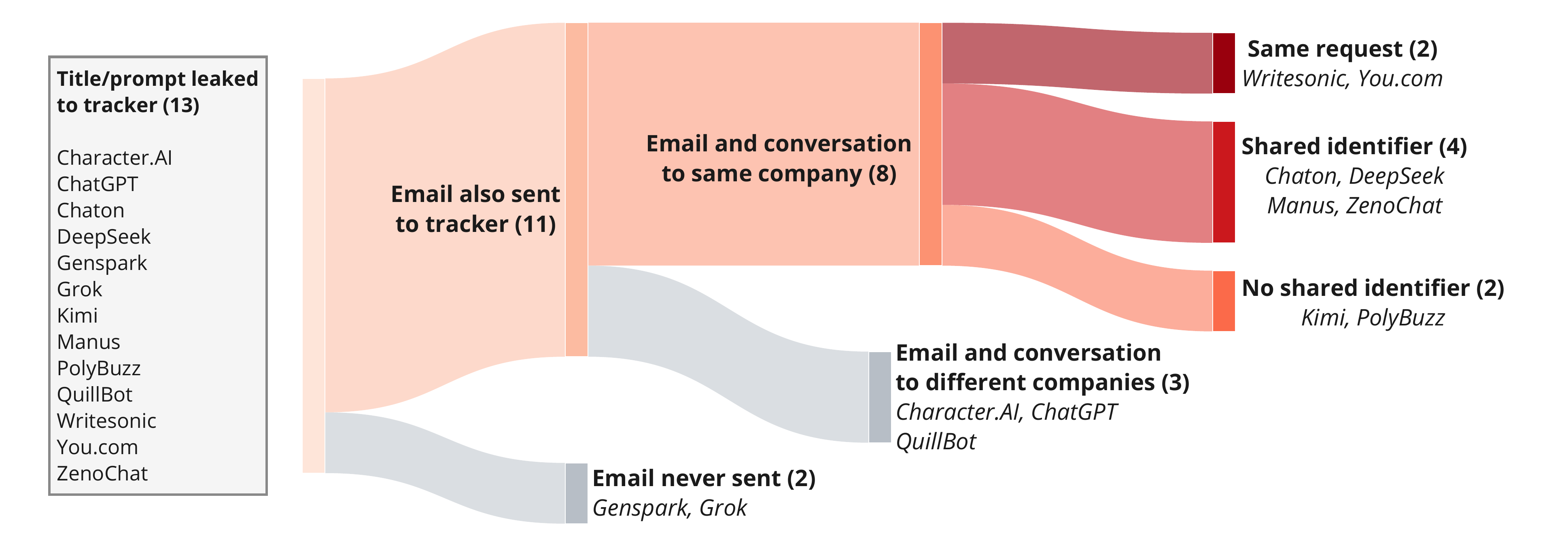}
  \caption{Identity correlation: 13 generic chatbot platforms leak the conversation to a
    third party, 11 also leak the email, and 8 of them send both to the same company  (6 with a directly observed linking identifier).}
  \label{fig:linkage-funnel}
\end{figure*}

Therefore, in this section, for each platform, we evaluate which of the 13 AI chatbot platforms that leak the title or the prompt of the conversation also share the registration email with a tracker. Then, we examine whether a \emph{single} tracker-owning company receives both these signals: the registration email (in raw, URL-encoded, or hashed form) and the leaked conversation context.
Furthermore, we check whether the recipient company can actually join the two, either because they arrive in the same outbound request, or by using a shared, persistent tracking cookie that ties the email and the context requests.
Our findings, for the 13 platforms that leak conversational context (the inferred title or the raw prompt), are reported in Figure~\ref{fig:linkage-funnel} and in Table~\ref{tab:linkage},

In the funnel diagram of Figure~\ref{fig:linkage-funnel}, we begin with the 13 platforms that leak the title or prompt on the left, and we progressively evaluate which of those also send the email identity to trackers.
Specifically, we observe that out of the 13 that make context-bearing requests to some tracker, 11 also send the user's email.
For eight of these, a single company receives \emph{both} the conversation and the email. For two of those eight chatbot platforms (Writesonic, You.com), the email and the conversational context can be found on \emph{the same outbound request}, so we observe the binding directly, while for four others, the respective company collects both signals, across separate requests which share the same unique identifier and can be associated.
Last, for Kimi and Polybuzz, we observe that both the email and the conversation reach the same company; however, this occurs over separate requests that bear no common token or identifier. Although we capture no shared token (cookie, URL parameter, etc.) between these requests, the same company still receives both requests which could be probabilistically linked together via the user's IP address or browser fingerprint.
The remaining five platforms either split the email and the conversation across different companies (Character.ai, ChatGPT, Quillbot), or never expose the email to any third party altogether (Genspark, Grok).

On Table~\ref{tab:linkage}, we attribute each aforementioned email-to-conversation linkage for the general-purpose AI chatbots to the tracker-owning company that receives the relevant requests. In this context, a single entity that receives both the email of a user and the conversational context of their interaction can associate the tracking data to the real user, thus fully identifying them. As such, we find that multiple AI chatbot platforms send both signals to more than one tracker-owning company (Writesonic, Chaton, Manus, ZenoChat), with Writesonic delivering both signals to six distinct companies.
In terms of the tracker-owning companies, we find that Google can re-identify users on at least four of the eight platforms, followed by Meta and ByteDance, mirroring the content-leakage concentration reported by Figure~\ref{fig:top_30_3p_domains_x_chatbots_request_counts}.

Overall, for eight of the thirteen context-leaking platforms, a single company can correlate the leaked conversation with the registration email, re-identifying the user and tying their chatbot interactions to their real-world identity. On four of these platforms, the exposure is distributed, as multiple tracker-owning companies receive both signals and can re-identify the user independently.

\subsection{Trackers in Task-specific AI Chatbots}
\label{subsection:findings_trackers_task_specific}

Having characterized general-purpose chatbot platforms, we now turn to the 17 task-specific chatbots across the education, health, and legal domains. In contrast to
general purpose AI chatbots, such platforms are purpose-built for contexts where submitted queries are inherently more likely to be sensitive in nature, making
contextual leakage to third-party trackers especially concerning.

\subsubsection{Tracker Landscape}
\label{subsubsection:findings_trackers_task_specific_tracker_landscape}

Across our interactions with the 17 task-specific platforms, we recorded
\totalTrackerRequests{} tracker-bound requests spanning \distinctTrackers{}
distinct tracker domains. On average, each prompt interaction generates
\allTotalRequests{} outbound requests, of which \allThirdPartyTrackerRequests{}
(\allTrackerPct\%) reach known trackers.
This tracker ratio is approximately uniform across the three chatbot categories: education
\eduThirdPartyTrackerRequests{}/\eduTotalRequests{} (\eduTrackerPct\%), health
\healthThirdPartyTrackerRequests{}/\healthTotalRequests{} (\healthTrackerPct\%), and
legal \legalThirdPartyTrackerRequests{}/\legalTotalRequests{} (\legalTrackerPct\%). It holds even though absolute volume varies sharply, with education platforms generating the most requests per prompt and health the fewest.
Notably, while task-specific AI platforms issue fewer total requests per prompt than general-purpose ones (\allTotalRequests{} vs.\
\statAllTotalReq{}), they devote a markedly larger fraction of that traffic to
trackers (\allTrackerPct\% vs.\ \statAllPctTracker\%). These smaller, single-purpose tools are proportionally more tracker-dense.

In terms of the tracker domains, traffic is concentrated towards the same companies that dominate the general-purpose ecosystem, as shown in the chatbot-to-owner Figure~\ref{fig:chatbots-tracker-owner-companies-task-specific} in the Appendix. Google is present on \emph{every} platform and, together with just Meta and Microsoft account for \emph{more than 70\%} of all tracker requests.
This reproduces the concentration we observed among general-purpose AI chatbots (Section~\ref{subsubsec:findings_trackers_generic_chatbots_tracker_traffic}), with the same handful of large advertising and analytics owners capturing the bulk of the tracker traffic.

\subsubsection{Context and Identity Leakage}
\label{subsubsection:findings_trackers_task_specific_context_and_identity_leakage}

Task-specific platforms leak conversational context as readily as general-purpose ones: 8 of the 17 send the conversation title to a tracker and five send the verbatim user prompt (Figures~\ref{fig:whole-string-title-chatbot-owner-task-specific}
and~\ref{fig:whole-string-prompt-chatbot-owner-task-specific} in the Appendix). Because these tools are purpose-built for sensitive domains, the leaked text is often deeply personal. For example, we observed that health chatbot platforms leak verbatim prompts such as ``I've been feeling really anxious and having trouble sleeping lately [..],'' while law-oriented chatbot platforms transmit prompts explicitly discussing the immigration status of the user. General-purpose chatbot leakage is already harmful, but in the case of task-specific chatbots, the sensitive conversational context that gets leaked carries medical and legal details, which users should be aware that they are sharing with an advertising network.

\begin{table}[t]
  \centering
  \setlength{\tabcolsep}{3.5pt}
  \renewcommand{\arraystretch}{1.1}
  \caption{Email and contextual leakage across general-purpose and task-specific chatbots.
    A single spanning entry means email and conversation reach the same company.
    Leak: T~=~title, P~=~prompt, E~=~email. For entries in the
    ``Different companies'' group, A~$/$~B denotes that A receives the email
    and B receives the conversation.}
  \label{tab:linkage}
  \begin{tabular}{@{}c l l p{4.5cm}@{}}
    \toprule
     & \textbf{Platform} & \textbf{Leak} & \textbf{Receiving Company}                                \\
    \midrule
    \multirow{4}{*}{\rotatebox[origin=c]{90}{\scriptsize\shortstack[c]{Same                          \\request}}}
     & Writesonic        & T, P, E       & Google, Meta, Microsoft, ByteDance, Mixpanel, Unifyintent \\
     & You.com           & T, P, E       & Google                                                    \\
     & AI Lawyer         & T, E          & Meta, ByteDance                                           \\
    \midrule
    \multirow{5}{*}{\rotatebox[origin=c]{90}{\scriptsize\shortstack[c]{Shared                        \\identifier}}}
     & Chaton            & T, E          & Google, Microsoft, Amplitude                              \\
     & DeepSeek          & T, E          & ByteDance                                                 \\
     & Manus             & T, E          & Meta, Amplitude                                           \\
     & ZenoChat          & T, P, E       & Google, Meta                                              \\
     & SciSpace          & T, P, E       & Google, Amplitude, Intercom                               \\
    \midrule
    \multirow{7}{*}{\rotatebox[origin=c]{90}{\scriptsize\shortstack[c]{No shared                     \\identifier}}}
     & Kimi              & T, P, E       & Google                                                    \\
     & PolyBuzz          & T, E          & Google                                                    \\
     & EduBrain          & T, P, E       & Google / Intercom                                         \\
     & Elicit            & T, E          & Google                                                    \\
     & iAsk              & T, P, E       & Google                                                    \\
     & iAsk Companion    & T, P, E       & Google                                                    \\
     & TheLawGPT         & T, E          & Google                                                    \\
    \midrule
    \multirow{4}{*}{\rotatebox[origin=c]{90}{\scriptsize\shortstack[c]{Different                     \\companies}}}
     & Character.AI      & T, E          & Google / Meta                                             \\
     & ChatGPT           & T, E          & Google / Meta                                             \\
     & QuillBot          & T, E          & Google, Microsoft / Meta                                  \\
     & AnswerAI          & T, P, E       & Google / Microsoft                                        \\
    \midrule
    \multirow{3}{*}{\rotatebox[origin=c]{90}{\scriptsize\shortstack[c]{Email                         \\not sent}}}
     & Genspark          & T             & Google, Meta, Microsoft, X~(Twitter), Naver               \\
     & Grok              & T             & Google                                                    \\
    \bottomrule
  \end{tabular}
\end{table}

Moreover, this content does not stay anonymous. On
Table~\ref{tab:linkage}, we present the context-leaking generic and
task-specific platforms together with the email leakage that accompanies the
conversation. For the task-specific chatbots, all eight platforms also transmit
the user's email to a third party, and for seven of them a single company
receives both these signals, allowing for a one-to-one linkage of the
conversation to the user's real-world identity. In the case of Edubrain,
Intercom receives the conversation but not the email of the user, while Google
receives the email and conversation context through different requests with no
shared identifier. As for the general-purpose, Google is the dominant recipient,
positioned to re-identify users on at least two of these seven platforms. In the
starkest case, the legal assistant AI Lawyer sends the conversation title and
the user's email in the \emph{same} request to Meta and ByteDance. As such, what
users treat as a personal health or legal consultation becomes an identified
record for advertisers.

\subsubsection{Leakage Across Task-Specific Chatbot Categories}
\label{subsubsection:findings_trackers_task_specific_leakage_across_categories}

\begin{figure}[t!]
  \centering
  \includegraphics[width=\columnwidth]{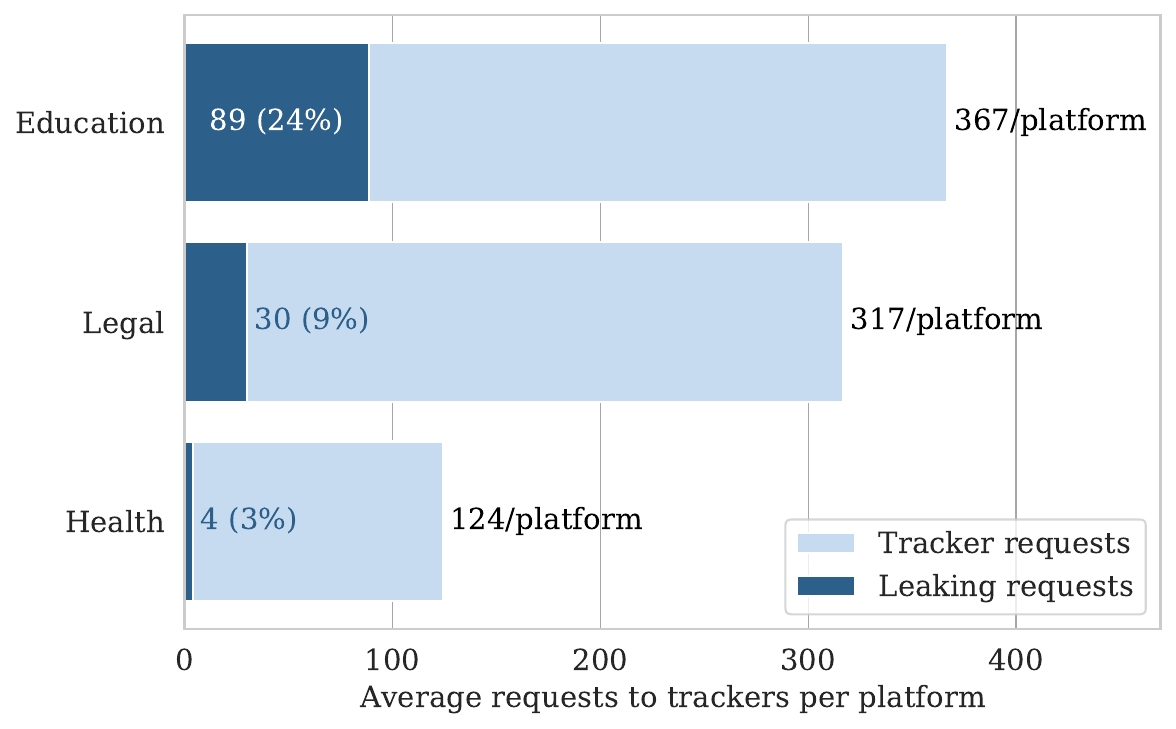}
  \caption{Per-category tracking and leakage: average tracker requests per platform (light), and the share that leak the title, prompt, or email (dark).}
  \label{fig:task-specific-tracker-leak-share}
\end{figure}

In Figure~\ref{fig:task-specific-tracker-leak-share}, we present the per-category average number of requests that each platform makes to trackers, 
along with the subset of those requests that carry the conversation title, prompt, or the user's email. We note that these category-level 
observations reflect the subset of platforms we were able to analyze within each category, rather than each category's ecosystem as a whole. 
Among the platforms we study, education platforms are both the heaviest trackers, at roughly 367 tracker requests per platform, and the most 
leak-prone, with about 24\% of those requests carrying user content. Legal platforms send a comparable volume (317 per platform) but leak a much 
smaller share (9\%), while health platforms track the least (124 per platform) and leak the least (3\%). Within each category the leakage is 
highly concentrated rather than uniform, e.g., SciSpace is responsible for roughly 80\% of education's leaking requests and AI Lawyer for nearly 
all of legal's, further indicating that these differences are driven by individual platforms rather than inherent to each category.

\subsection{Interaction Decomposition}
\label{subsection:interaction_decomp}

\begin{figure}[t!]
  \centering
  \includegraphics[width=\columnwidth]{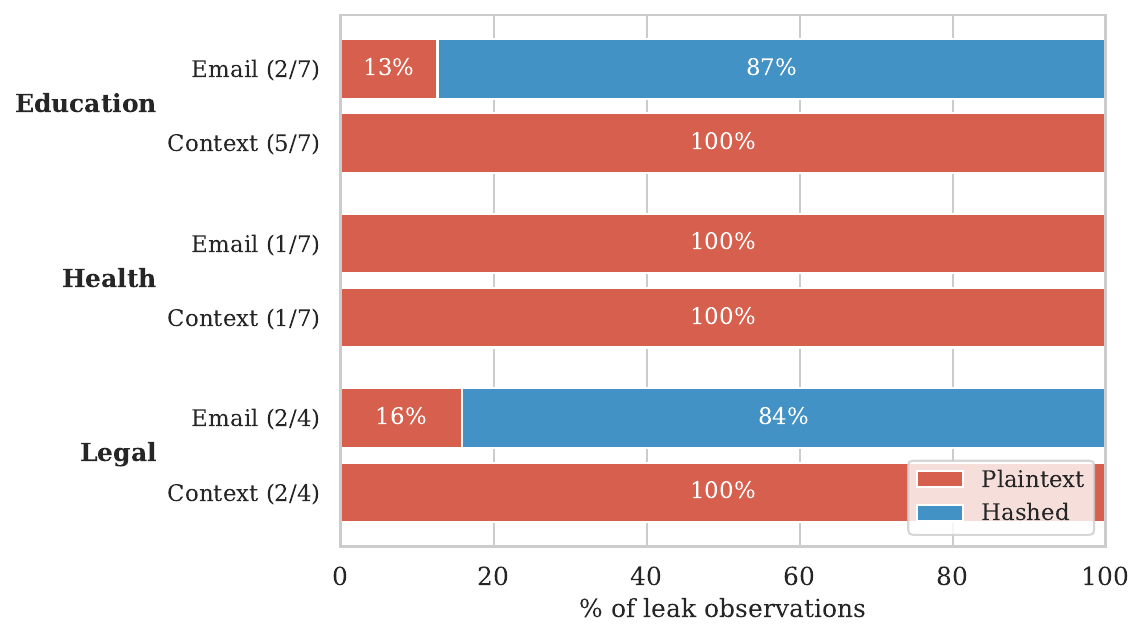}
  \caption{Email vs.\ content leak encoding by category. The \emph{Email} and \emph{Content} rows show how each category encodes the leaked email and the leaked title/prompt (plaintext vs.\ hashed); the counts give the number of platforms per category (of its total) that leak.}
  \label{fig:task-specific-email-vs-content-form}
\end{figure}

Across all three categories, a substantial share of each platform's traffic reaches trackers, even though the fraction that carries leaked context differs, 
with the education platforms in our set leaking the most and the health platforms the least. It is already concerning that these single-purpose tools leak 
sensitive conversations to trackers at all, and more concerning that they do so at this scale. Moreover, we observe that a smaller leakage footprint does 
not necessarily translate into better protections for the users. In Figure~\ref{fig:task-specific-email-vs-content-form}, we present how each category encodes the email and the conversation content when they leak. 
Across all three categories, the conversation content, whether the title or the prompt, is transmitted in plaintext. In contrast, the email presents 
a less uniform approach, with the education and legal platforms predominantly hashing it and sending a SHA-256 digest of the email address. 
However, the health platform that leaks the email transmits it in completely unencrypted form. Although the health chatbots in our set represent 
the smallest leakage footprint, sending the email in raw form represents a more direct privacy risk. Unlike hashed identifiers, a raw email address 
requires no additional processing to link a user to their interactions, making it straightforward to re-identify the user. As such, even a platform 
with a small leakage footprint can pose a significant identity-privacy risk, handing third parties the data needed to deanonymize the user in plaintext.

To determine which user actions trigger contextual leakage, we isolate five interaction scenarios per platform, each executed in a separate VV8 session
with independent execution traces. For this experiment, we focused only on the platforms where we found contextual leakage to third parties.
From our analysis we exclude two platforms (ChatGPT and Grok) which have updated their web interface and removed the existing trackers before our
experiment took place. Additionally, we also exclude two character-based AI chatbots where the title of the conversation is directly leaked upon
navigation and corresponds to the character the user interacts with. For these two platforms we did not find any third party tracker collecting the full conversation context
(prompt). We start by analyzing all outbound network operations for the presence of two markers: the raw user prompt and the platform-generated
conversation title. By comparing the third-party destinations observed in each scenario against the prompt submission baseline, we identify which user
action is the causal trigger for each leakage flow.

Table~\ref{tab:scenario-leakage} summarizes which platforms exhibit contextual leakage to at least one third party under each scenario. Six platforms
already leak conversational context upon \emph{submission} of the prompt, before any navigation takes place. Specifically, on Zenochat and Writesonic, Microsoft Clarity captures the raw prompt
text from the input text field in the DOM even before the platform has generated the title of the conversation. On You.com the prompt is embedded directly on the
page URL so every tracker that populates their payload with the \texttt{location.href} (googleads.g.doubleclick.net, www.google.com, www.googleadservices.com)
receives the whole prompt upon submission found in the \texttt{tiba, url, ref} query parameters. On Kimi, Genspark and DeepSeek, the platform sets the conversation title fast
enough that it is already present in \texttt{document.title} when the first analytics page view fires during the initial session.

Because conversation titles are dynamically generated upon response, some tracking scripts have fired the corresponding event before the title is
rendered in the DOM. This is evident as we see that \emph{all} nine platforms expose the conversation title to third parties during navigation to an
existent conversation. This is because navigating within a single-page application, which handles different chats as sub-pages, fires a synthetic
page view event in embedded analytics and advertising tags, which read and transmit the sub-page title, which also holds the title of the conversation.

The remaining scenarios produce minimal additional leakage. \emph{Scroll} and \emph{Click} re-fire the same analytics tags already observed on Submit and Navigate,
without introducing new third-party destinations. \emph{Delete} triggers leakage on only one platform (Writesonic), where deleting a conversation causes
a page view event that carries the title to Google Analytics. These results indicate that the primary leakage vectors are the initial prompt submission
and the first subsequent navigation, with other interactions contributing little beyond what these two scenarios already expose.

\begin{table}[t!]
  \centering
  \caption{Prompt or title leakage to third parties by interaction.}
  \label{tab:scenario-leakage}
  \begin{tabular}{|l|c|c|c|c|c|}
    \hline
    \textbf{Platform} & \textbf{Submit}              & \textbf{Navigate}            & \textbf{Delete}              & \textbf{Scroll}              & \textbf{Click}               \\
    \hline
    Zenochat          & \cellcolor{red!30}\checkmark & \cellcolor{red!30}\checkmark & \cellcolor{blue!20}\ding{55} & \cellcolor{red!30}\checkmark & \cellcolor{red!30}\checkmark \\
    \hline
    You.com           & \cellcolor{red!30}\checkmark & \cellcolor{red!30}\checkmark & \cellcolor{blue!20}\ding{55} & \cellcolor{red!30}\checkmark & \cellcolor{red!30}\checkmark \\
    \hline
    Quillbot          & \cellcolor{blue!20}\ding{55} & \cellcolor{red!30}\checkmark & \cellcolor{blue!20}\ding{55} & \cellcolor{blue!20}\ding{55} & \cellcolor{blue!20}\ding{55} \\
    \hline
    Manus             & \cellcolor{blue!20}\ding{55} & \cellcolor{red!30}\checkmark & \cellcolor{blue!20}\ding{55} & \cellcolor{blue!20}\ding{55} & \cellcolor{blue!20}\ding{55} \\
    \hline
    Kimi              & \cellcolor{red!30}\checkmark & \cellcolor{red!30}\checkmark & \cellcolor{blue!20}\ding{55} & \cellcolor{red!30}\checkmark & \cellcolor{red!30}\checkmark \\
    \hline
    DeepSeek          & \cellcolor{red!30}\checkmark & \cellcolor{red!30}\checkmark & \cellcolor{blue!20}\ding{55} & \cellcolor{blue!20}\ding{55} & \cellcolor{blue!20}\ding{55} \\
    \hline
    Genspark          & \cellcolor{red!30}\checkmark & \cellcolor{red!30}\checkmark & \cellcolor{blue!20}\ding{55} & \cellcolor{red!30}\checkmark & \cellcolor{red!30}\checkmark \\
    \hline
    Writesonic        & \cellcolor{red!30}\checkmark & \cellcolor{red!30}\checkmark & \cellcolor{red!30}\checkmark & N/A                          & N/A                          \\
    \hline
    ChatOn            & \cellcolor{blue!20}\ding{55} & \cellcolor{red!30}\checkmark & \cellcolor{blue!20}\ding{55} & N/A                          & N/A                          \\
    \hline
  \end{tabular}
\end{table}

\begin{table*}[t!]
  \centering
  \caption{VV8 contextual capture mechanisms used by trackers on AI chat platforms.}
  \label{tab:capture-mechanisms}
  \scalebox{0.85}{
    \begin{tabular}{llll}
      \toprule
      \textbf{Capture Method}                  & \textbf{Operation Description}             & \textbf{Trackers}                            & \textbf{Platforms}              \\
      \midrule
      \texttt{document.title} (getter)         & Property access on \texttt{document.title} & Google Analytics, Clarity, Naver, X/Twitter, & Zenochat, Writesonic, ChatOn,    \\
                                               &                                            & Amplitude, Volcengine, TikTok Pixel          & Genspark, Kimi, Manus, Quillbot, \\
                                               &                                            &                                              & DeepSeek \\
      \addlinespace
      \texttt{document.querySelector("title")} & DOM query for \texttt{<title>} element,    & Meta Pixel, TikTok Pixel                     & Zenochat, Genspark, Manus,      \\
                                               & then reads \texttt{textContent}            &                                              & Quillbot, Writesonic             \\
      \addlinespace
      \texttt{document.getElementsByTagName}   & DOM query for \texttt{<title>}             & Google Analytics                             & Zenochat                        \\
      \texttt{("title")}                       & element collection                         &                                              &                                 \\
      \addlinespace
      \texttt{element.textContent} /           & Walks DOM tree, reads all visible          & Clarity                                      & Zenochat, Writesonic,            \\
      \texttt{element.value}                   & text and form values                       &                                              & Quillbot                        \\
      \addlinespace
      \texttt{element.getAttribute} +          & Reads HTML attributes and text of          & Amplitude, Sentry                            & Manus                           \\
      \texttt{element.textContent}             & clicked/interacted DOM elements            &                                              &                                 \\
      \addlinespace
      \texttt{location.href}                   & Reads page URL which contains              & Google Ads                                   & You.com                         \\
                                               & prompt directly                            &                                              &                                 \\
      \bottomrule
    \end{tabular}}
\end{table*}

While these isolated scenarios allow us to infer \emph{when} contextual leakage occurs, we cannot explain \emph{how} each tracker obtains the conversation
context. However, VV8's execution traces record every JavaScript API call a script performs allowing us to link each outbound request to the specific read
operation that preceded it. Analyzing these traces, starting from the outbound requests containing the taint values (prompts and titles) we identify the execution
script, resolve its origin and enumerate the read operations performed on the DOM before transmission. In our analysis, we correlate reads to transport via
script identifier: if the same script context initiates a request towards a third party endpoint and also called \texttt{document.title} as a property getter, we consider that a read and transmission chain.

As illustrated by Table~\ref{tab:capture-mechanisms}, the most prevalent mechanism is a property read of \texttt{document.title} observed on eight out of
the nine platforms. Analytics services such as Google Analytics and Volcengine, as well as session replay tools like Clarity, utilize this standard property
to access the specific value and populate the corresponding parameter (e.g., \texttt{dt=} or \texttt{tl=}) values for outbound requests.

On the other hand, Meta Pixel employs an entirely different approach.
On all five platforms where it is present (Zenochat, Genspark, Manus, Quillbot, and Writesonic),
its script (\texttt{fbevents.js}) never accesses the \texttt{document.title} property. Instead, it calls \texttt{document.querySelector("title")} to select the
\texttt{<title>} HTML element from the DOM and then reads its \texttt{textContent} to obtain the page title. Interestingly, TikTok Pixel on Writesonic uses both mechanisms: it reads \texttt{document.title}
as a property and also queries the DOM via \texttt{querySelector}.

Microsoft Clarity, present on Zenochat, Writesonic, and Quillbot, captures context through a broader mechanism. Beyond reading \texttt{document.title},
Clarity walks the DOM tree and reads \texttt{textContent} and \texttt{value} properties across page elements as part of its session replay functionality.
Thus, Clarity captures not only the conversation title but also visible conversation text rendered in the page, including the \emph{raw prompt} even before
a title has been generated.

On Manus, we observe a capture mechanism unique among the platforms we study. Amplitude's autocapture feature records every user click, logging the
\texttt{textContent} and HTML attributes of the clicked element and its ancestors. When a user clicks a conversation entry in the sidebar, 
Amplitude captures the \texttt{title} attribute of the \texttt{<span>} element along with its visible text. Similarly, Sentry's performance 
monitoring records the CSS selector of UI interactions, which includes the element's \texttt{title} attribute.

These mechanisms, deployed by third-party tracking frameworks, were designed for ``traditional'' websites. Page titles are static descriptors and rarely carry
sensitive information. On the other hand, on the web interfaces of AI chatbots, the design pattern of treating each conversation as a navigable sub-page with a
dynamically generated, context-derived title that inadvertently transforms these mechanisms to capture and propagate conversational context. Finally, as also observed by the outbound requests, You.com represents a distinct leakage behavior. The platform embeds the user's prompt directly
in the initial page URL so any tracker that reads the \texttt{location.href} receives the prompt without needing to access the page title at all. Google Ads operating
on You.com obtains the prompt through this mechanism, in the values of the \texttt{tiba, url} and \texttt{ref} parameters.

\subsection{Privacy Policies}

Finally, we look further into the Privacy and Cookie Policies~\cite{characterai-privacy-policies, chatgpt-privacy-policies, chaton-privacy-policies, deepseek-privacy-policies, genspark-privacy-policies, grok-privacy-policies, kimi-privacy-policies, manus-privacy-policies, polybuzz-privacy-policies, quilbot-privacy-policies, writesonic-privacy-policies, you-privacy-policies, zenochat-privacy-policies, ailawyer-privacy-policies, answerai-privacy-policies, edubrain-privacy-policies, elicit-privacy-policies, scispace-privacy-policies, thelawgpt-privacy-policies, iask-privacy-policies}
of the platforms where we identified contextual leakage. Our findings suggest a more subtle and systemic issue: AI chatbot platforms integrate third-party analytics,
advertising, and session replay frameworks under the same assumptions that govern their deployment on traditional websites. In multiple instances we
find that the privacy notices state that such frameworks are deployed to collect information about the browsing activity of the user within the platform
covering page-views, click events, or session duration.

Notably, several platforms explicitly state in their privacy notices that user data is not sold or shared for targeted advertising purposes, yet we
observe third-party advertising and analytics pixels actively operating on their interfaces and collecting conversational context. This discrepancy
suggests that platforms may not fully account for the data collection practices of the third-party frameworks they integrate,
beyond what is explicitly configured. Third-party tracking pixels and similar analytics frameworks are designed to also capture structural page components
in order to measure engagement and attribute conversions. On a conventional e-commerce website for example, such signals carry little
sensitive information: a page titled "Men's Running Shoes" or a URL containing \texttt{/checkout} reveals commercial intent.

AI chatbot platforms, however, fundamentally alter this assumption. Unlike static web pages, chatbot interfaces dynamically construct some of their
structural components from user input. The prompt itself reveals the user's intent, topic of interest, or personal circumstances directly, and conversation
titles are auto-generated semantic compressions of the user's prompt that are set as the \emph{title} of the current chat session.
Each chat instance is treated as a separate sub-page having a unique title which however encapsulates conversation context. As described
in Table~\ref{tab:capture-mechanisms}, the same capture mechanisms that trackers are implemented to capture page components describing navigation patterns
and commercial interactions now are retrieving conversational context rather than generic page metadata. This structural shift raises significant privacy concerns
as we find that privacy policies disclose the operation of these pixels covering pageviews and interaction metrics, with no acknowledgment that on a chatbot,
these signals carry a categorically different informational payload.

  \subsection*{Key Takeaways}
  \begin{itemize}[leftmargin=15pt, topsep=2pt, itemsep=8pt, parsep=0pt]
    \item \emph{Pervasive third-party tracking.} AI chatbot platforms embed conventional third-party tracking and analytics frameworks, which are the same as those deployed across
          traditional websites.
    \item \emph{Contextual leakage is systemic.} We audit 42 platforms and
    find that 21 of them leak conversation titles and nine leak verbatim user prompts
    to third-party trackers, indicating that contextual leakage is a systemic
    issue.
    \item \emph{Leaked conversational context is re-identifiable.} Trackers receiving conversational context frequently also obtain the user's email, allowing even sensitive disclosures to be tied to a named individual.
    \item \emph{Conventional analytics is a privacy risk on AI chatbots.} Tracking tools on AI chatbot platforms require careful scrutiny: deploying traditional-web analytics on AI chatbots should be treated as a privacy risk, since such tools capture conversational context that users reasonably expect to remain private.
  \end{itemize}

\section{Limitations}
\label{sec:limitations}

\subsubsection*{Experimental Repetitions} Each platform was measured under identical conditions and interaction scenarios, 
varying only the registration email, between May 6 and May 20, 2026. Our results describe the state of these platforms during that period. 
Tracker deployments change as platforms add or remove vendors, so the set of platforms leaking conversational context, and the third parties
receiving it, may differ outside our measurement window.

\subsubsection*{Location and Consent} We measured from a single US-East vantage point with tracking consent accepted. This configuration isolates 
our central question, namely whether contextual leakage occurs and through what mechanism. Consent state and region determine which trackers a platform loads, 
so we do not study whether these platforms honor consent choices or how their deployments vary across jurisdictions.

\subsubsection*{Platform Coverage} We study web-based chatbot platforms that offer free account registration, which excludes paid tiers and the growing 
number of AI chatbots distributed primarily as mobile or desktop applications. Extending this analysis to application-based clients would provide additional
insights regarding the state of this ecosystem.

\section{Related Work}
\label{section:llm_leaks_related_work}

\subsubsection*{Model and Application Leakage}
A separate line of work has studied the privacy issues and leakage vectors originating from the LLM itself or from applications built on top of it.
At the model level, studies have focused on
personally identifiable information
(PII) memorization in pretrained models~\cite{kim2023propile,zanellabeguelin2020updates}, multi-step jailbreaks
extracting personal information from ChatGPT~\cite{li2023multistep} or inappropriate information disclosure~\cite{mireshghallah2024confaide}.
At the application level, Priyanshu\etal~\cite{priyanshu2023regurgitation} documented prompt regurgitation in chatbots, while Yan\etal~\cite{yan2025file}
studied how file exfiltration content could be achieved through prompt injection in third-party GPT-based applications. Moreover, Shakeel \etal~\cite{shakeel2025investigating}
demonstrated that PII retained across sessions by AI chatbots can be exploited to disclose a user's personal information in unrelated conversations or to adversarial
actors. These studies address privacy risks inherent to the LLM or its application layer but do not consider the tracking infrastructure embedded in the chatbots' web
interfaces, where third-party scripts can silently extract conversation context and share it with third-party actors.

\subsubsection*{Third-party Tracking on the Web}
Privacy in the online ecosystem has been extensively studied with works focused on identifying and measuring the prevalence
of third-party actors on the web, both through client-side~\cite{fbclid-paper,inv-pixels,cookie-swap,cookie-graph,saly,mayer2012thirdparty,pii-leakage-4}
and server-side~\cite{meta-server-side,server-side} tracking mechanisms. Studies have also examined the presence and impact of such trackers
on ``sensitive'' websites~\cite{sick-and-tracked,k12-schools,religious,third-party-cookies-hospitals,medical},  highlighting the increased concern when
tracking occurs on platforms that handle sensitive user information. For
example, Moti \etal~\cite{children-websites-tracking} measured tracking and
advertising components on
children-related websites, while Huo \etal~\cite{phi-leakage} explored the
disclosure of protected health information to third parties from within
web-based healthcare systems.

Complementing these measurement efforts, numerous studies have characterized the
PII that trackers
collect~\cite{pii-leakage-1, pii-leakage-2, pii-leakage-3, pii-leakage-4,
    contact-forms-leakage, leaky-forms,
    not-register}.  Kubicek
\etal~\cite{registration-forms-large-scale} automated the submission of
registration forms to study PII collection by analyzing outbound traffic towards
third-party destinations. Kieserman \etal~\cite{kieserman2025tracker} measured such PII leakage by injecting 
artificial forms into websites and inspecting the network traffic generated upon 
their submission. More recently, studies have turned to analyzing the
structural components of tracking implementations themselves, uncovering
\emph{contextual} leakage as an additional privacy risk. Bekos
\etal~\cite{piixel-leaks} demonstrated that Meta Pixel can collect
contextual information describing user interactions with a website's components,
while Ghani \etal~\cite{passive-3} further examined the structural configuration
of Meta Pixel deployments that enable such collection practices.

\subsubsection*{Third-party Tracking on AI Chatbots}
Although AI chatbots have become an integral part of online activity, the presence of third-party trackers on their web interfaces has received limited attention.
Existing work has primarily focused on the ChatGPT ecosystem, studying how third-party developers build plugins and apps that interface with the platform and the
data collection practices these integrations introduce~\cite{gpt-specific-1,gpt-specific-2}. Vekaria \etal~\cite{bigbrother-extension} audited nine GenAI browser
extensions for tracking, profiling, and personalization, revealing that extensions can exfiltrate sensitive page content, including form inputs. Concurrently to
our work, Jazlan \etal~\cite{jazlan2026tracking} measured content and identity exposure across 20 general-purpose AI chatbots by capturing network traffic during
a single-prompt session. Similarly, Girish \etal~\cite{leakylm} examined third-party tracking across four chatbot platforms through HAR traffic analysis,
documenting the transmission of conversation URLs and user identifiers to advertising and analytics services.

Our work fundamentally differs from prior studies in that we treat tracking as an \emph{interaction-driven} phenomenon. Tracking pixels and analytics frameworks are fundamentally
event-driven, designed to trigger on user interactions rather than solely on page load. A single-prompt network capture is therefore insufficient to understand when
and how contextual leakage occurs on these platforms. The class-action complaints recently filed against both OpenAI~\cite{lim-v-openai,couture-v-openai} and
Perplexity~\cite{doe-v-perplexity} further underscore this point, alleging that embedded trackers transmitted conversational context to third parties upon user
\emph{interaction} with the chatbot.
Our study addresses these gaps by capturing both contextual and identity leakage towards third-party actors across structured multi-interaction
sessions, and then decomposing these sessions into isolated interaction scenarios to attribute each data flow to its causal trigger. Our analysis through VV8 
further shows that trackers capture this context through unmodified, generic page-metadata primitives, a structural finding that network-only measurements 
cannot establish.

\section{Discussion}
\label{section:discussion}

Third-party tracking scripts have been designed to operate on traditional websites, where the structural components they capture, such as page titles and URLs,
describe generic user interactions like product views or general content engagement. However, when deployed on AI chatbot platforms under the same default configuration,
these same mechanisms retrieve structural components that encode conversational context: the user's prompt and the dynamically generated conversation
title derived from it. This distinction is fundamental, as the metadata of a chat session do not include only generic page information but also a direct reflection of
what the user has disclosed. Deploying such frameworks on AI chatbot platforms therefore requires deliberate consideration of the privacy implications of the
page's structural components, and these platforms should not be treated as traditional websites when configuring third-party tracking integrations.

Our findings reveal a systematic mismatch between the intent of conventional tracking frameworks and the context in which they are deployed on AI chatbot platforms. 
Standard client-side privacy tools remain relevant here, as blocking these scripts would prevent the observed leakage. A more principled long-term 
direction, however, is the development of AI-specific analytics frameworks that decouple usage measurement from conversational content, allowing 
operators to collect meaningful platform metrics without exposing user intent to third parties. Such an outcome could be approached from two directions. Platform developers could 
minimize the embedding of conversational context in dynamic page components, for example by generating less specific conversation titles that omit 
sensitive details rather than directly reflecting the user's query. Third-party vendors, on the other hand, could develop dedicated analytics solutions for conversational interfaces, or 
extend existing frameworks with stricter privacy constraints that prevent the collection of content-bearing page attributes in chatbot contexts.

Furthermore, when conducting privacy audits of such platforms, developers should evaluate the full range of user interactions rather than limiting their assessment to page
load or prompt submission events. Tracking pixels are event-driven by nature and may fire on interactions such as navigation, button clicks, or chat
deletion, each of which can independently trigger contextual leakage that would otherwise go undetected.

During the course of our study, ChatGPT and Grok had third party trackers operating on their platforms that were collecting conversational context.
Both platforms have since removed these trackers, which we verified prior to submission, suggesting that the issue is actionable and that platforms
can remediate it once identified. For all remaining platforms where contextual leakage was observed, we have followed responsible disclosure practices
and notified the respective teams. During our responsible disclosure process, four platforms (Manus, Kimi, QuillBot and SciSpace) received our notification and 
corresponding data and forwarded them to the appropriate teams, two platforms (Genspark and Chaton) acknowledged receipt of the disclosure, 
and one platform (TheLawGPT) directly remediated the issue by updating the configuration of the corresponding tracker responsible for the leakage.

\section{Conclusion}

In this study, we present a systematic measurement study of contextual leakage in AI chatbot platforms, auditing 42 platforms spanning
general-purpose and task-specific chatbots in sensitive domains. We find that contextual leakage is a systemic phenomenon, with 21 platforms
transmitting conversation titles and 9 transmitting verbatim user prompts to third-party trackers. Beyond content leakage, the majority of
context-leaking platforms also expose the user's email to the same trackers, enabling direct linkage of conversational context to a named
individual. Through VV8-based interaction decomposition, we attribute each leakage flow to its causal trigger and reconstruct the
JavaScript-level mechanisms through which trackers extract conversational context from the page.

Our findings highlight a fundamental mismatch between how AI chatbot platforms are instrumented and the nature of the content they handle.
Unlike traditional websites, chatbot interfaces dynamically encode user intent into structural page components, causing standard tracking integrations
to inadvertently capture even \emph{sensitive} user disclosures. Therefore, these platforms should be carefully evaluated for third-party tracking deployments in this
context and privacy audits must cover the full range of user interactions, as leakage may manifest under different interaction events.

\section{Acknowledgments}

We would like to thank the anonymous reviewers for their constructive comments and feedback. This work was supported by 
the National Science Foundation (NSF) under grant CNS-2211575,
with additional support
by Amazon through the Amazon Research Awards program.
Any opinions, findings, and conclusions or recommendations
expressed in this material are those of the authors, and do not
necessarily reflect the views of the funding agencies and sponsoring
organizations
that supported this work.

\bibliographystyle{IEEEtran}
\bibliography{bibliography}

\begin{thebibliography}{100}
\providecommand{\url}[1]{#1}
\csname url@samestyle\endcsname
\providecommand{\newblock}{\relax}
\providecommand{\bibinfo}[2]{#2}
\providecommand{\BIBentrySTDinterwordspacing}{\spaceskip=0pt\relax}
\providecommand{\BIBentryALTinterwordstretchfactor}{4}
\providecommand{\BIBentryALTinterwordspacing}{\spaceskip=\fontdimen2\font plus
\BIBentryALTinterwordstretchfactor\fontdimen3\font minus
  \fontdimen4\font\relax}
\providecommand{\BIBforeignlanguage}[2]{{%
\expandafter\ifx\csname l@#1\endcsname\relax
\typeout{** WARNING: IEEEtran.bst: No hyphenation pattern has been}%
\typeout{** loaded for the language `#1'. Using the pattern for}%
\typeout{** the default language instead.}%
\else
\language=\csname l@#1\endcsname
\fi
#2}}
\providecommand{\BIBdecl}{\relax}
\BIBdecl

\bibitem{chatbot-statistics-1}
{Backlinko}, ``{ChatGPT / OpenAI Statistics: How Many People Use ChatGPT?}''
  \url{https://backlinko.com/chatgpt-stats}.

\bibitem{chatbot-statistics-2}
{Instant Press}, ``{AI \& ChatGPT Statistics for 2026},''
  \url{https://www.instantpress.co/ai-statistics}.

\bibitem{chatbot-statistics-3}
{Elfsight}, ``{Chatbot Statistics 2025–2026: Market Size, Usage \& Trends},''
  \url{https://elfsight.com/blog/chatbot-statistics-and-trends/}.

\bibitem{chatbot-statistics-4}
{Brookings}, ``{How are Americans using AI? Evidence from a nationwide
  survey},''
  \url{https://www.brookings.edu/articles/how-are-americans-using-ai-evidence-from-a-nationwide-survey/}.

\bibitem{fbclid-paper}
P.~Bekos, P.~Papadopoulos, E.~P. Markatos, and N.~Kourtellis, ``The
  hitchhiker’s guide to {Facebook} web tracking with invisible pixels and
  click {IDs},'' in \emph{Proceedings of the 32nd ACM Web Conference (WWW)},
  2023.

\bibitem{inv-pixels}
I.~Fouad, N.~Bielova, A.~Legout, and N.~Sarafijanovic-Djukic, ``Missed by
  filter lists: Detecting unknown third-party trackers with invisible pixels,''
  in \emph{Proceedings of the Privacy Enhancing Technologies Symposium (PETS)},
  2020.

\bibitem{cookie-swap}
Q.~Chen, P.~Ilia, M.~Polychronakis, and A.~Kapravelos, ``Cookie swap party:
  Abusing first-party cookies for web tracking,'' in \emph{Proceedings of the
  30th ACM Web Conference (WWW)}, 2021.

\bibitem{cookie-graph}
S.~Munir, S.~Siby, U.~Iqbal, S.~Englehardt, Z.~Shafiq, and C.~Troncoso,
  ``{CookieGraph}: Understanding and detecting first-party tracking cookies,''
  in \emph{Proceedings of the ACM SIGSAC Conference on Computer and
  Communications Security (CCS)}, 2023.

\bibitem{saly}
S.~Dambra, I.~Sanchez-Rola, L.~Bilge, and D.~Balzarotti, ``When {Sally} met
  trackers: Web tracking from the users' perspective,'' in \emph{Proceedings of
  the 31st USENIX Security Symposium}, 2022.

\bibitem{children-websites-tracking}
Z.~Moti, A.~Senol, H.~Bostani, F.~Z. Borgesius, V.~Moonsamy, A.~Mathur, and
  G.~Acar, ``Targeted and troublesome: Tracking and advertising on children’s
  websites,'' in \emph{Proceedings of the IEEE Symposium on Security and
  Privacy (S\&P)}, 2024.

\bibitem{sick-and-tracked}
X.~Yu, N.~Samarasinghe, M.~Mannan, and A.~Youssef, ``Got sick and tracked:
  Privacy analysis of hospital websites,'' in \emph{Proceedings of the IEEE
  European Symposium on Security and Privacy Workshops (EuroS\&PW)}, 2022.

\bibitem{k12-schools}
J.~Chanenson, B.~Sloane, N.~Rajan, A.~Morril, J.~Chee, D.~Y. Huang, and
  M.~Chetty, ``Uncovering privacy and security challenges in {K-12} schools,''
  in \emph{Proceedings of the Conference on Human Factors in Computing Systems
  (CHI)}, 2023.

\bibitem{religious}
N.~Samarasinghe, P.~Kapoor, M.~Mannan, and A.~Youssef, ``No salvation from
  trackers: Privacy analysis of religious websites and mobile apps,'' in
  \emph{Proceedings of the Data Privacy Management, Cryptocurrencies and
  Blockchain Technology: ESORICS International Workshops}, 2023.

\bibitem{phi-leakage}
M.~Huo, M.~Bland, and K.~Levchenko, ``All eyes on me: Inside third party
  trackers' exfiltration of {PHI} from healthcare providers' online systems,''
  in \emph{Proceedings of the 21st Workshop on Privacy in the Electronic
  Society}, 2022.

\bibitem{medical}
S.~Rauti, E.~Vuorinen, R.~Carlsson, and P.~Puhtila, ``Data leaks to third-party
  services on medical websites,'' in \emph{Proceedings of the 16th
  International Conference on Security of Information and Networks (SIN)},
  2023.

\bibitem{lim-v-openai}
``Lim v. {OpenAI} global, {LLC},'' May 2026.

\bibitem{couture-v-openai}
``Couture v. {OpenAI} global, {LLC},'' May 2026.

\bibitem{doe-v-perplexity}
``Doe v. {Perplexity AI}, inc., {Meta} platforms, inc., and {Google} {LLC},''
  Mar. 2026.

\bibitem{jazlan2026tracking}
M.~Jazlan, E.~Wang, Y.~Vekaria, and Z.~Shafiq, ``Tracking conversations:
  Measuring content and identity exposure on ai chatbots,'' 2026.

\bibitem{leakylm}
A.~Girish, G.~Oliveira, G.~Suarez-Tangil, J.~G. Herrero, M.~Sanchez,
  N.~Vallina-Rodriguez, and T.~Jackevicius, ``{LeakyLM: AI Assistants Are
  Leaking Your Conversations},'' \url{https://leakylm.github.io/}, 2026.

\bibitem{vv8}
J.~Jueckstock and A.~Kapravelos, ``{VisibleV8}: In-browser monitoring of
  javascript in the wild,'' in \emph{Proceedings of the Internet Measurement
  Conference (IMC)}, 2019.

\bibitem{gpt-specific-1}
C.~Yan, R.~Ren, M.~H. Meng, L.~Wan, T.~Y. Ooi, and G.~Bai, ``Exploring
  {ChatGPT} app ecosystem: Distribution, deployment and security,'' in
  \emph{Proceedings of the 39th IEEE/ACM International Conference on Automated
  Software Engineering (ASE)}, 2024.

\bibitem{gpt-specific-2}
Y.~Wu, E.~Jaff, K.~Yang, N.~Zhang, and U.~Iqbal, ``An in-depth investigation of
  data collection in {LLM} app ecosystems,'' in \emph{Proceedings of the ACM
  Internet Measurement Conference (IMC)}, 2025.

\bibitem{koi-report}
{KOI Research}, ``{8 Million Users' AI Conversations Sold for Profit by
  ``Privacy'' Extensions},''
  \url{https://www.koi.ai/blog/urban-vpn-browser-extension-ai-conversations-data-collection}.

\bibitem{piixel-leaks}
P.~Bekos, P.~Papadopoulos, N.~Kourtellis, and M.~Polychronakis, ``{PIIxel
  Leaks}: Passive identification of personally identifiable information leakage
  through {Meta Pixel},'' in \emph{Proceedings of the ACM SIGSAC Conference on
  Computer and Communications Security (CCS)}, 2025.

\bibitem{passive-3}
A.~Ghani, Y.~Vekaria, and Z.~Shafiq, ``{PixelConfig}: Longitudinal measurement
  and reverse-engineering of {Meta Pixel} configurations,'' in
  \emph{Proceedings of the ACM Internet Measurement Conference (IMC)}, 2026.

\bibitem{pii-leakage-1}
H.~Dao and K.~Fukuda, ``Alternative to third-party cookies: Investigating
  persistent {PII} leakage-based web tracking,'' in \emph{Proceedings of the
  17th International Conference on Emerging Networking EXperiments and
  Technologies (CoNEXT)}, 2021.

\bibitem{pii-leakage-2}
J.~Brookman, P.~Rouge, A.~Alva, and C.~Yeung, ``Cross-device tracking:
  Measurement and disclosures,'' in \emph{Proceedings of the Privacy Enhancing
  Technologies Symposium (PETS)}, 2017.

\bibitem{pii-leakage-3}
S.~Englehardt, J.~Han, and A.~Narayanan, ``I never signed up for this!
  {Privacy} implications of email tracking,'' in \emph{Proceedings of the
  Privacy Enhancing Technologies Symposium (PETS)}, 2018.

\bibitem{pii-leakage-4}
G.~Acar, S.~Englehardt, and A.~Narayanan, ``No boundaries: Data exfiltration by
  third parties embedded on web pages,'' in \emph{Proceedings of the Privacy
  Enhancing Technologies Symposium (PETS)}, 2020.

\bibitem{contact-forms-leakage}
O.~Starov, P.~Gill, and N.~Nikiforakis, ``Are you sure you want to contact us?
  {Quantifying} the leakage of {PII} via website contact forms,'' in
  \emph{Proceedings of the Privacy Enhancing Technologies Symposium (PETS)},
  2016.

\bibitem{leaky-forms}
A.~Senol, G.~Acar, M.~Humbert, and F.~Z. Borgesius, ``Leaky forms: A study of
  email and password exfiltration before form submission,'' in
  \emph{Proceedings of the 31st USENIX Security Symposium}, 2022.

\bibitem{registration-forms-large-scale}
K.~Kubicek, J.~Merane, A.~Bouhoula, and D.~Basin, ``Automating website
  registration for studying {GDPR} compliance,'' in \emph{Proceedings of the
  ACM Web Conference (WWW)}, 2024.

\bibitem{not-register}
M.~Chatzimpyrros, K.~Solomos, and S.~Ioannidis, ``You shall not register!
  {D}etecting privacy leaks across registration forms,'' in \emph{Computer
  Security}.\hskip 1em plus 0.5em minus 0.4em\relax Springer-Verlag, 2019.

\bibitem{kieserman2025tracker}
J.~B. Kieserman, A.~Andreou, C.~Geeng, T.~Lauinger, and D.~McCoy, ``Tracker
  installations are not created equal: Understanding tracker configuration of
  form data collection,'' in \emph{Proceedings of the Privacy Enhancing
  Technologies Symposium (PETS)}, 2025.

\bibitem{similar-web}
``Top ai chatbots and tools websites ranking,''
  \url{https://www.similarweb.com/top-websites/ai-chatbots-and-tools/}, 2026.

\bibitem{copilot}
{Microsoft}, ``{Microsoft Copilot},'' \url{https://copilot.microsoft.com/},
  2025.

\bibitem{deepseek}
{DeepSeek}, ``{DeepSeek},'' \url{https://chat.deepseek.com/}, 2025.

\bibitem{claude}
{Anthropic}, ``{Claude AI},'' \url{https://claude.ai/}, 2025.

\bibitem{mistrallechat}
{Mistral AI}, ``{Le Chat Mistral},'' \url{https://chat.mistral.ai/chat}.

\bibitem{zenochat}
{TextCortex AI}, ``{Zenochat},'' \url{https://textcortex.com/}.

\bibitem{metaai}
{Meta}, ``{Meta AI},'' \url{https://www.meta.ai/}, 2025.

\bibitem{characterai}
{Google}, ``{character.ai},'' \url{https://character.ai/}.

\bibitem{quilbot}
{Learneo}, ``{QuillBot},'' \url{https://quillbot.com/}.

\bibitem{manus}
{Manus}, ``{manus},'' \url{https://manus.im/}.

\bibitem{huggingchat}
{Hugging Face}, ``{Hugging Chat},'' \url{https://huggingface.co/chat/}.

\bibitem{qwen}
{Alibaba Cloud}, ``{Qwen},'' \url{https://qwen.ai/home}.

\bibitem{grok}
{xAI}, ``{Grok},'' \url{https://grok.com/}, 2025.

\bibitem{chatsonic}
{Writesonic}, ``{Writesonic},'' \url{https://app.writesonic.com/}.

\bibitem{you}
{SuSea, Inc.}, ``{you.com},'' \url{https://you.com/chat}.

\bibitem{kimi}
{Moonshot AI}, ``{Kimi AI},'' \url{https://www.kimi.com/}.

\bibitem{perplexity}
{Perplexity AI}, ``{Perplexity},'' \url{https://www.perplexity.ai/}, 2025.

\bibitem{chatgpt}
{OpenAI}, ``{ChatGPT},'' \url{https://chatgpt.com/}, 2025.

\bibitem{poe}
{Quora}, ``{Poe},'' \url{https://poe.com/}, 2025.

\bibitem{duckduckgo}
{Duck Duck Go, Inc.}, ``{Duck.ai},'' \url{https://duck.ai/}.

\bibitem{seart}
{SUNRISEAI PTE. LTD}, ``{SEAART AI},'' \url{https://www.seaart.ai/}.

\bibitem{pi}
{Inflection AI}, ``{Pi},'' \url{https://pi.ai}.

\bibitem{genspark}
{Mainfunc Inc.}, ``{Genspark AI Workspace},'' \url{https://www.genspark.ai/}.

\bibitem{chaton}
{AIBY}, ``{ChatOn},'' \url{https://chaton.ai/}.

\bibitem{polybuzz}
{Cloud Whale Interactive Technology LLC}, ``{PolyBuzz},''
  \url{https://www.polybuzz.ai/}.

\bibitem{gemini}
{Google}, ``{Gemini},'' \url{https://gemini.google.com/app}, 2025.

\bibitem{khanmigo}
{Khan Academy}, ``{Khanmigo},'' \url{https://www.khanmigo.ai/}.

\bibitem{edubrain}
{EduBrain}, ``{EduBrain},'' \url{https://edubrain.ai/}.

\bibitem{flexi}
{CK-12 Foundation}, ``{Flexi},'' \url{https://www.ck12.org/flexi/}.

\bibitem{scispace}
{SciSpace}, ``{SciSpace},'' \url{https://scispace.com/}.

\bibitem{elicit}
{Elicit AI}, ``{Elicit},'' \url{https://elicit.com/}.

\bibitem{iaskai}
{AI Search Inc.}, ``{iAsk AI},'' \url{https://iask.ai/}.

\bibitem{answerai}
{Answer.AI}, ``{Answer.AI},'' \url{https://answerai.pro/}.

\bibitem{freeainotatherapist}
{Free AI (not a) Therapist}, ``{Free AI (not a) Therapist},''
  \url{https://www.freeainotatherapist.com/}.

\bibitem{feelbetterbot}
{FeelBetterBot}, ``{FeelBetterBot},'' \url{https://feelbetterbot.com/}.

\bibitem{wellzy}
{Wellzy Inc.}, ``{Wellzy},'' \url{https://wellzy.io/}.

\bibitem{iaskai-companion}
{AI Search Inc.}, ``{iAsk AI - Companion Mode},''
  \url{https://iask.ai/companion/}.

\bibitem{earkick}
{Earkick AI}, ``{Earkick},'' \url{https://chat.earkick.com/}.

\bibitem{abby}
{Abby Intelligence, Inc.}, ``{Abby},'' \url{https://abby.gg/}.

\bibitem{aitherapistfree}
{AITherapistFree}, ``{AITherapistFree},''
  \url{https://www.aitherapistfree.com/}.

\bibitem{lawconnect}
{LawConnect}, ``{LawConnect},'' \url{https://lawconnect.com/}.

\bibitem{nyayguru}
{NyayGuru AI}, ``{NyayGuru},'' \url{https://nyayguru.com/}.

\bibitem{ailawyer}
{AI Lawyer}, ``{AI Lawyer},'' \url{https://app.ailawyer.pro/}.

\bibitem{thelawgpt}
{TheLawGPT}, ``{TheLawGPT},'' \url{https://www.thelawgpt.com/}.

\bibitem{puppeteer}
``Puppeteer,'' \url{https://pptr.dev/}, 2018.

\bibitem{disconnect-list}
{Disconnect}, ``{Tracker Protection Lists},''
  \url{https://github.com/disconnectme/disconnect-tracking-protection}.

\bibitem{tracker-radar}
{DuckDuckGo}, ``{DuckDuckGo Tracker Radar},''
  \url{https://github.com/duckduckgo/tracker-radar}.

\bibitem{characterai-privacy-policies}
{Character.AI}, ``{Character.AI Privacy Policy},''
  \url{https://policies.character.ai/privacy}, 2025.

\bibitem{chatgpt-privacy-policies}
{OpenAI}, ``{Privacy policy},''
  \url{https://openai.com/policies/row-privacy-policy/}, 2025.

\bibitem{chaton-privacy-policies}
{Chaton}, ``{Privacy Policy},'' \url{https://chaton.ai/legal/privacy/en/},
  2025.

\bibitem{deepseek-privacy-policies}
{Deepseek}, ``{DeepSeek Privacy Policy},''
  \url{https://cdn.deepseek.com/policies/en-US/deepseek-privacy-policy.html},
  2025.

\bibitem{genspark-privacy-policies}
{Genspark}, ``{Privacy Policy},'' \url{https://www.genspark.ai/privacy}, 2025.

\bibitem{grok-privacy-policies}
{xAI}, ``{xAI Privacy Policy},'' \url{https://x.ai/legal/privacy-policy}, 2025.

\bibitem{kimi-privacy-policies}
{Moonshot AI PTE. LTD}, ``{Privacy Policy},''
  \url{https://www.kimi.com/user/agreement/userPrivacy?version=v2}, 2025.

\bibitem{manus-privacy-policies}
{Butterfly Effect Pte. Ltd}, ``{Privacy},'' \url{https://manus.im/privacy},
  2025.

\bibitem{polybuzz-privacy-policies}
{Polybuzz}, ``{Privacy Policy},'' \url{https://www.polybuzz.ai/privacy-policy},
  2025.

\bibitem{quilbot-privacy-policies}
{Quillbot}, ``{Quillbot's Privacy Policy},''
  \url{https://quillbot.com/privacy}, 2025.

\bibitem{writesonic-privacy-policies}
{Writesonic}, ``{Privacy Policy},''
  \url{https://writesonic.com/legal/privacy-policy}, 2025.

\bibitem{you-privacy-policies}
{You.com}, ``{Privacy Policy},'' \url{https://you.com/privacy}, 2025.

\bibitem{zenochat-privacy-policies}
{TextCortex}, ``{TextCortex Privacy Policy},''
  \url{https://textcortex.com/privacy-policy}, 2025.

\bibitem{ailawyer-privacy-policies}
{AI Lawyer}, ``{Privacy Policy},'' \url{https://ailawyer.pro/privacy-policy},
  2025.

\bibitem{answerai-privacy-policies}
{Answer.AI}, ``{Answer.AI Privacy Policy},''
  \url{https://answerai.pro/privacy-policy.html}, 2025.

\bibitem{edubrain-privacy-policies}
{EduBrain}, ``{Privacy Policy},'' \url{https://edubrain.ai/privacy-policy/},
  2025.

\bibitem{elicit-privacy-policies}
{Elicit AI}, ``{Privacy Policy},'' \url{https://elicit.com/operations/privacy},
  2025.

\bibitem{scispace-privacy-policies}
{SciSpace}, ``{Privacy},'' \url{https://scispace.com/privacy}, 2025.

\bibitem{thelawgpt-privacy-policies}
{TheLawGPT}, ``{Privacy Policy},''
  \url{https://www.thelawgpt.com/privacy-policy}, 2025.

\bibitem{iask-privacy-policies}
{AI Search Inc.}, ``{Privacy Policy},'' \url{https://iask.ai/privacy}, 2025.

\bibitem{kim2023propile}
S.~Kim, S.~Yun, H.~Lee, M.~Gubri, S.~Yoon, and S.~J. Oh, ``{ProPILE}: Probing
  privacy leakage in large language models,'' in \emph{Advances in Neural
  Information Processing Systems (NeurIPS)}, 2023.

\bibitem{zanellabeguelin2020updates}
S.~Zanella-B{\'e}guelin, L.~Wutschitz, S.~Tople, V.~R{\"u}hle, A.~Paverd,
  O.~Ohrimenko, B.~K{\"o}pf, and M.~Brockschmidt, ``Analyzing information
  leakage of updates to natural language models,'' in \emph{Proceedings of the
  ACM SIGSAC Conference on Computer and Communications Security (CCS)}, 2020.

\bibitem{li2023multistep}
H.~Li, D.~Guo, W.~Fan, M.~Xu, J.~Huang, F.~Meng, and Y.~Song, ``Multi-step
  jailbreaking privacy attacks on {ChatGPT},'' in \emph{Findings of the
  Association for Computational Linguistics (EMNLP)}, 2023.

\bibitem{mireshghallah2024confaide}
N.~Mireshghallah, H.~Kim, X.~Zhou, Y.~Tsvetkov, M.~Sap, R.~Shokri, and Y.~Choi,
  ``Can {LLMs} keep a secret? {T}esting privacy implications of language models
  via contextual integrity theory,'' in \emph{Proceedings of the International
  Conference on Learning Representations (ICLR)}, 2024.

\bibitem{priyanshu2023regurgitation}
A.~Priyanshu, S.~Vijay, A.~Kumar, R.~Naidu, and F.~Mireshghallah, ``Are
  chatbots ready for privacy-sensitive applications? {An} investigation into
  input regurgitation and prompt-induced sanitization,'' 2023.

\bibitem{yan2025file}
C.~Yan, B.~Guan, Y.~Li, M.~H. Meng, L.~Wan, and G.~Bai, ``Understanding and
  detecting file knowledge leakage in {GPT} app ecosystem,'' in
  \emph{Proceedings of the ACM Web Conference (WWW)}, 2025.

\bibitem{shakeel2025investigating}
A.~Shakeel and S.~Thaseen, ``Investigating leakage of sensitive information in
  {LLM}-powered chatbots,'' in \emph{Proceedings of the International
  Conference on Computational Intelligence and Knowledge Economy (ICCIKE)},
  2025.

\bibitem{mayer2012thirdparty}
J.~R. Mayer and J.~C. Mitchell, ``Third-party web tracking: Policy and
  technology,'' in \emph{Proceedings of the IEEE Symposium on Security and
  Privacy (S\&P)}, 2012.

\bibitem{meta-server-side}
A.~El~fraihi, N.~Amieur, W.~Rudametkin, and O.~Goga, ``Client-side and
  server-side tracking on {Meta}: Effectiveness and accuracy,'' in
  \emph{Proceedings on Privacy Enhancing Technologies Symposium (PETS)}, 2024.

\bibitem{server-side}
I.~Fouad, C.~Santos, and P.~Laperdrix, ``The devil is in the details:
  Detection, measurement and lawfulness of server-side tracking on the web,''
  in \emph{Proceedings on Privacy Enhancing Technologies Symposium (PETS)},
  2024.

\bibitem{third-party-cookies-hospitals}
A.~B. Friedman, R.~M. Merchant, A.~Maley, K.~Farhat, K.~Smith, J.~Felkins,
  R.~E. Gonzales, L.~Bauer, and M.~S. McCoy, ``Widespread third-party tracking
  on hospital websites poses privacy risks for patients and legal liability for
  hospitals,'' \emph{Health Affairs (Millwood)}, 2023.

\bibitem{bigbrother-extension}
Y.~Vekaria, A.~L. Canino, J.~Levitsky, A.~Ciechonski, P.~Callejo, A.~M.
  Mandalari, and Z.~Shafiq, ``Big help or big brother? {Auditing} tracking,
  profiling, and personalization in generative {AI} assistants,'' in
  \emph{Proceedings of the 34th USENIX Security Symposium}, 2025.

\end{thebibliography}
\appendices

\section{Extended Findings}

\begin{figure}[h]
  \centering
  \includegraphics[width=0.9\linewidth]{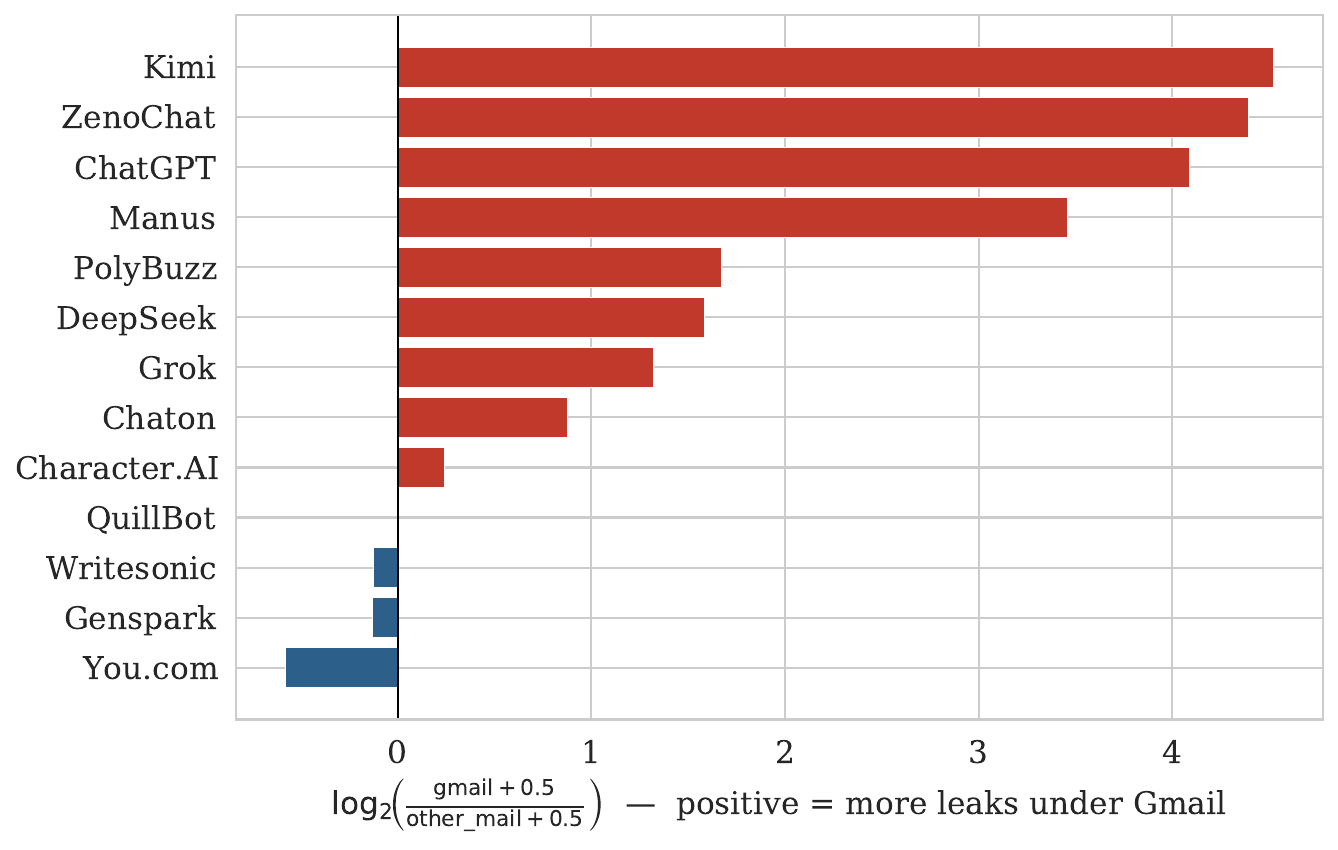}
  \caption{}
  \label{fig:mail_provider_asymmetry_of_whole_string_leak_volume}
\end{figure}

Figure~\ref{fig:mail_provider_asymmetry_of_whole_string_leak_volume} presents the email--provider asymmetry of whole-string (title or prompt) leak volume, 
measured using the per-platform $\log_2$ ratio of leaks  under a Gmail vs.\ non-Gmail account, with one session per configuration. Positive means more leaks under Gmail. 
Kimi, ZenoChat, ChatGPT, and Manus leak $\sim$12--24$\times$ more under Gmail,
with only You.com, Writesonic and Genspark leaning the other way.

\begin{figure}[t]
  \centering
  \includegraphics[width=\linewidth]{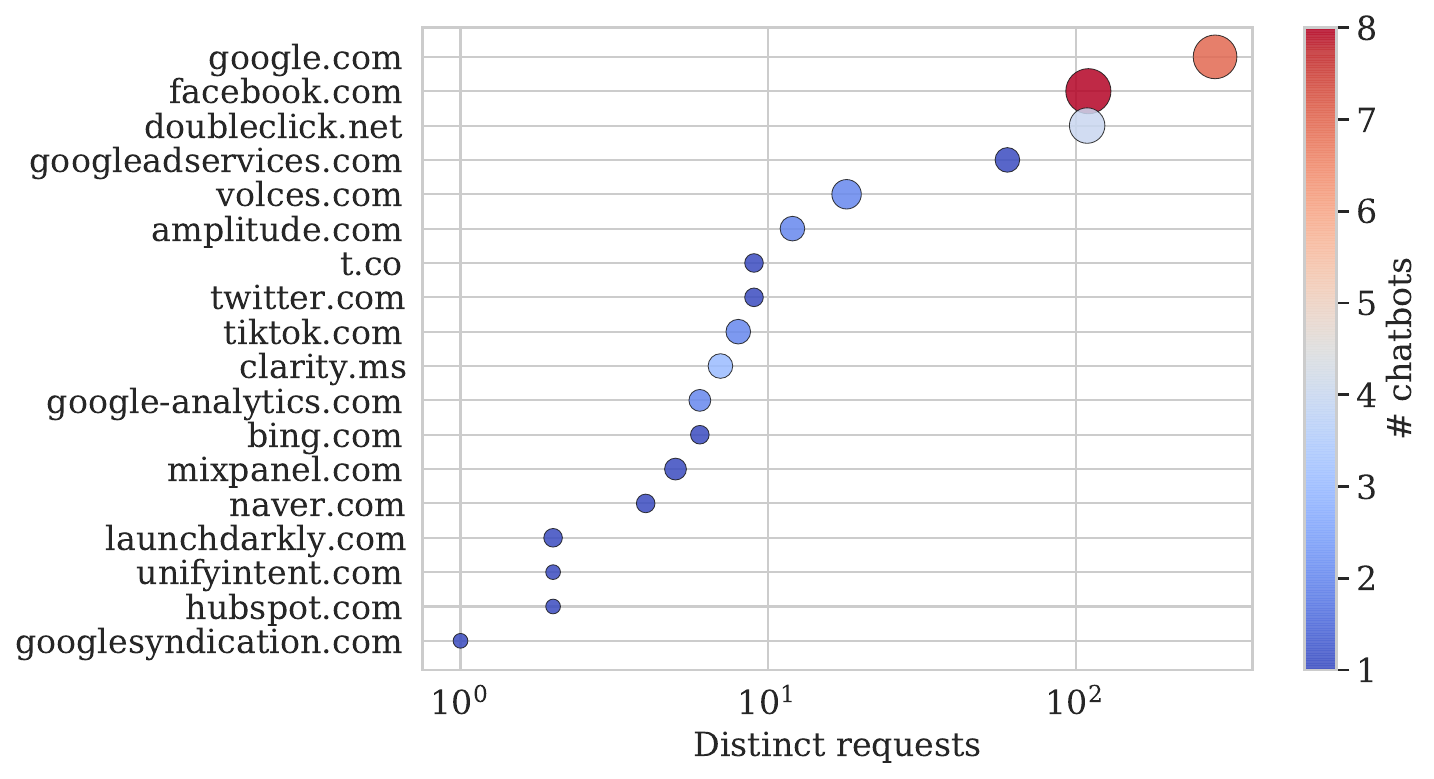}
  \caption{Distinct context-bearing requests to tracker domains.}
  \label{fig:whole_string_leaks_top_3p_domains_distinct_requests}
\end{figure}

Figure~\ref{fig:whole_string_leaks_top_3p_domains_distinct_requests} provides a per-domain breakdown of the data aggregated by company in Figure~\ref{fig:whole_string_leaks_top_3p_domains_distinct_requests_grouped_by_owners}.
Google's dominance is distributed across multiple distinct endpoints, with \emph{google.com}, \emph{doubleclick.net}, and \emph{googleadservices.com} each
independently accumulating substantial context-bearing requests, underscoring
the breadth of its tracking infrastructure across the chatbot ecosystem.

\begin{figure}[t]
  \centering
  \includegraphics[width=\columnwidth]{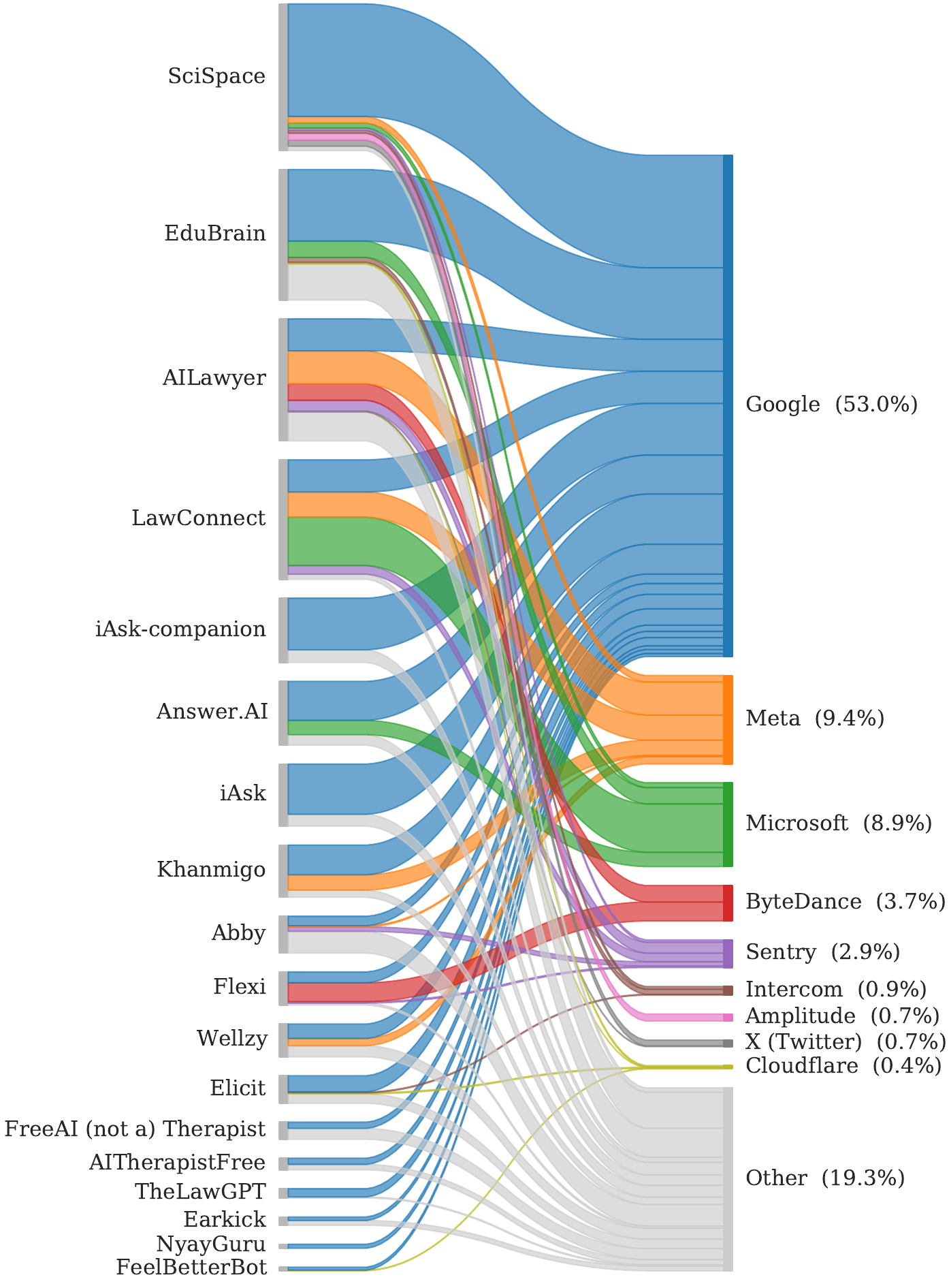}
  \caption{Flow of third-party requests from task-specific chatbots to tracker owner companies.}
  \label{fig:chatbots-tracker-owner-companies-task-specific}
\end{figure}

Similarly to our findings for the generic AI chatbots, we find that the ecosystem of third-party companies operating on these platforms is densely populated
by three companies: Google, Meta and Microsoft, as shown in Figure~\ref{fig:chatbots-tracker-owner-companies-task-specific}. These companies account for 70\% of all tracking-related requests. More importantly, we find that all
chatbot platforms heavily favour the adoption of Google services, as it is the company found in all the network traces.

\begin{figure}[t]
  \centering
  \includegraphics[width=\columnwidth]{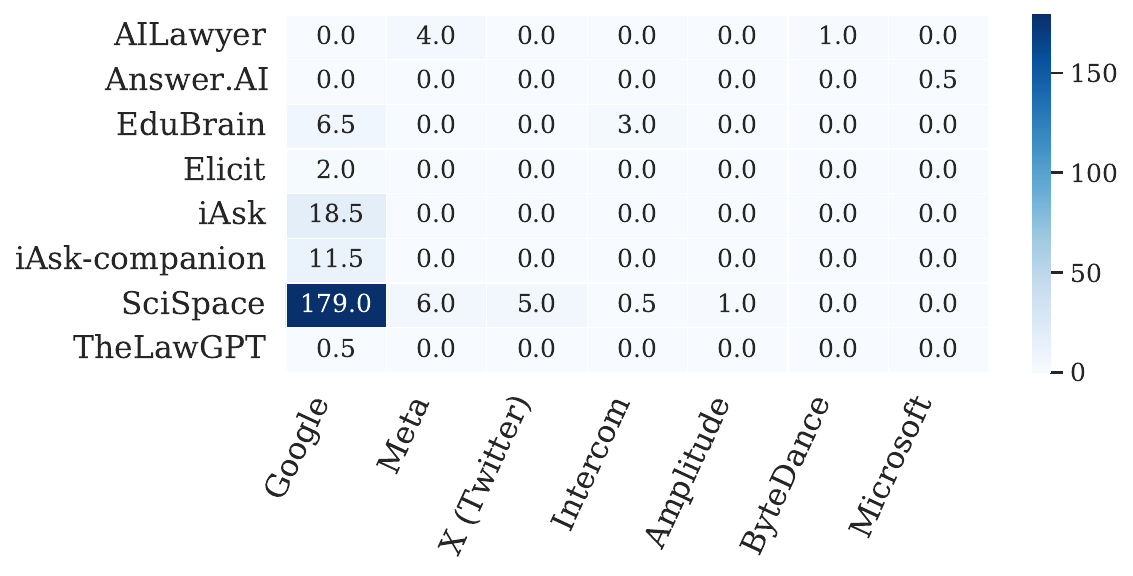}
  \caption{Whole-string title leaks per task-specific chatbot $\times$ owner company, averaged per prompt.}
  \label{fig:whole-string-title-chatbot-owner-task-specific}
\end{figure}

\begin{figure}[t]
  \centering
  \includegraphics[width=\columnwidth]{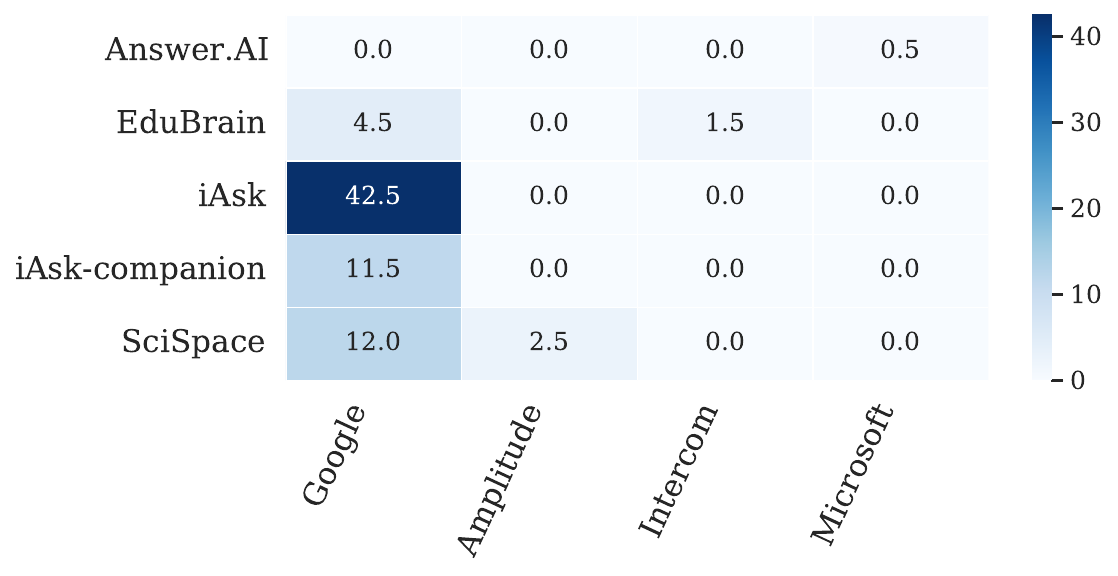}
  \caption{Whole-string prompt leaks per task-specific chatbot $\times$ owner company, averaged per prompt.}
  \label{fig:whole-string-prompt-chatbot-owner-task-specific}
\end{figure}

As illustrated by Figure~\ref{fig:whole-string-title-chatbot-owner-task-specific}, we find eight task-specific chatbots sharing conventional context,
in the form of \emph{titles} with a total of seven tracker-owning companies. Interestingly enough, similarly to You.com, we find a heavy concentration
of third party requests that include conversation titles towards Google from
SciSpace. In general, we find heavier conventional leakage patterns in
academic and educational
chatbots. Finally, as illustrated by Figure~\ref{fig:whole-string-prompt-chatbot-owner-task-specific}, we also find five instances of full \emph{prompt}
leakage towards four companies operating third-party analytics and tracking scripts on these platforms.

\section{Open Science}
We provide an artifact accompanying this paper at \url{https://zenodo.org/records/22831655}.
It includes (i) the network traffic logs collected across all 42 platforms and (ii) the VisibleV8 JavaScript execution traces,
each accompanied by the analysis scripts that reproduce our findings.

\section{AI Usage}

In this paper, we conduct a passive measurement study on the privacy implications of interacting with online AI chatbot platforms. As such,
AI platforms are the object of the study. Specifically, we audit 42 general-purpose and task-specific chatbots for third-party leakage of conversational
context by passively measuring the requests that these platforms send during ordinary use. In terms of using generative AI as a tool in the development
of this work, we employed Anthropic's Claude Opus 4.7 to generate the prompts that were used in the interactions with the AI chatbot platforms,
and to develop and refine our data analysis pipeline. All generative AI outputs were manually verified and meticulously reviewed. Generative AI was used
for editorial purposes in this manuscript, and all outputs were inspected by the authors to ensure accuracy and originality. Specifically, we used
Claude to tighten and spellcheck the manuscript, as well as to fix grammar and syntax errors. As above, all editorial generative AI suggestions were
inspected by the authors and manually verified.

\section{Ethics Considerations}
Our study involves passive measurement of publicly accessible AI chatbot web interfaces. No real user data was collected or processed at any point,
as our experiments were conducted using two dedicated accounts created exclusively for this study, with no involvement of real users or their data.
Interactions with live platforms were conducted in a controlled, low-volume manner using browser instrumentation (Puppeteer and VisibleV8) across manual
sessions. We generated a minimal set of prompts per platform and a minimal number of interactions per session, deliberately avoiding traffic patterns
that could disrupt normal platform operations. We do not access, modify, exfiltrate, or interfere with other users' data or platform functionality.
Our methodology is strictly observational, recording only outbound network traffic and JavaScript execution traces originating from our own controlled
sessions.

\newpage
\section{Meta-Review}
The following meta-review was prepared by the program committee for the 2027
IEEE Symposium on Security and Privacy (S\&P) as part of the review process as
detailed in the call for papers.

\subsection{Summary}

This paper presents a measurement study of privacy leakage in the web interfaces of AI chatbot platforms, spanning both general-purpose and 
domain-specific systems. Using a combination of Puppeteer-based network interception and VisibleV8 JavaScript instrumentation, 
the authors analyze how conversational data and user identifiers are transmitted to third-party trackers, attribute each leak 
to specific user interactions, and identify the underlying JavaScript mechanisms responsible. Their analysis reveals that leakage of 
conversation titles and verbatim user prompts are present and that a substantial amount of chatbot platforms leak users' registration 
email addresses, enabling the association of sensitive conversational content with user identities. The study further shows that many 
of these leaks arise unintentionally through generic metadata collection, such as the extraction of dynamically generated page titles 
that contain conversation summaries. The paper highlights widespread privacy risks in current AI chatbot deployments and provides 
new insights into the interactions and implementation mechanisms that lead to contextual leakage.

\subsection{Scientific Contributions}
\begin{itemize}
\item Identifies an Impactful Vulnerability
\item Provides a Valuable Step Forward in an Established Field
\end{itemize}

\subsection{Reasons for Acceptance}
\begin{enumerate}
\item The paper identifies an important privacy vulnerability in the web interfaces of AI chatbots specifically studying health, legal, and education platforms 
and providing valuable insight into the privacy implications of sensitive-domain applications.
\item This paper provides a valuable step forward in an established field. Web privacy is an ongoing concern due to the widespread collection 
and tracking of user data. The paper presents novel characterization of contextual leakage and explains when and how it occurs in the context of 
AI chatbots. By attributing leaks to specific user interactions and identifying the DOM-read primitives used by individual trackers, the 
paper provides insights that network-only analyses cannot.
\end{enumerate}

\subsection{Noteworthy Concerns}

Consistency and reproducibility of the measured behavior: Reviewers raised concerns the paper relies on a limited number of measurements and sessions.

\end{document}